\input harvmac.tex
 \input epsf.tex

\def\figin{\epsfcheck\figin}\def\figins{\epsfcheck\figins}
\def\epsfcheck{\ifx\epsfbox\UnDeFiNeD
\message{(NO epsf.tex, FIGURES WILL BE IGNORED)}
\gdef\figin##1{\vskip2in}\gdef\figins##1{\hskip.5in}
\else\message{(FIGURES WILL BE INCLUDED)}%
\gdef\figin##1{##1}\gdef\figins##1{##1}\fi}
\def\DefWarn#1{}
\def\figinsert{\goodbreak\midinsert}
\def\ifig#1#2#3{\DefWarn#1\xdef#1{fig.~\the\figno}
\writedef{#1\leftbracket fig.\noexpand~\the\figno} %
\figinsert\figin{\centerline{#3}}\medskip\centerline{\vbox{\baselineskip12pt
\advance\hsize by -1truein\noindent\footnotefont{\bf
Fig.~\the\figno:} #2}}
\bigskip\endinsert\global\advance\figno by1}



\def \la {\langle}
\def \ra {\rangle}
\def \beq  {\begin{eqnarray}}
\def \eeq  {\end{eqnarray}}
\def \pa {\partial}

\def \eps {\epsilon}


\def\frac#1#2{{#1 \over #2}}
\def\text#1{#1}

\def\Ward{charge conservation }


\lref\FitzpatrickZM{
  A.~L.~Fitzpatrick, E.~Katz, D.~Poland and D.~Simmons-Duffin,
JHEP\ {\bf 1107}, 023  (2011).
[arXiv:1007.2412 [hep-th]].
}
\lref\BelavinVU{
  A.~A.~Belavin, A.~M.~Polyakov, A.~B.~Zamolodchikov,
Nucl.\ Phys.\  {\bf B241}, 333-380 (1984).
}
\lref\SagnottiAT{
  A.~Sagnotti and M.~Taronna,
Nucl.\ Phys.\ B\ {\bf 842}, 299  (2011).
[arXiv:1006.5242 [hep-th]].
}

\lref\ColemanAD{
  S.~R.~Coleman, J.~Mandula,
Phys.\ Rev.\  {\bf 159}, 1251-1256 (1967).
}

\lref\ZamolodchikovXM{
  A.~B.~Zamolodchikov, A.~B.~Zamolodchikov,
Annals Phys.\  {\bf 120}, 253-291 (1979).
}
\lref\ShenkerZF{
  S.~H.~Shenker and X.~Yin,
[arXiv:1109.3519 [hep-th]].
}
\lref\KlebanovJA{
  I.~R.~Klebanov, A.~M.~Polyakov,
Phys.\ Lett.\  {\bf B550}, 213-219 (2002).
[hep-th/0210114].
}
\lref\FradkinQY{
  E.~S.~Fradkin and M.~A.~Vasiliev,
  Nucl.\ Phys.\  B {\bf 291}, 141 (1987).
}
\lref\GiombiRZ{
  S.~Giombi, S.~Prakash, X.~Yin,
[arXiv:1104.4317 [hep-th]].
}

\lref\MikhailovBP{
  A.~Mikhailov,
[hep-th/0201019].
}

\lref\DouglasRC{
  M.~R.~Douglas, L.~Mazzucato and S.~S.~Razamat,
Phys.\ Rev.\ D\ {\bf 83}, 071701  (2011).
[arXiv:1011.4926 [hep-th]].
}

\lref\CostaMG{
  M.~S.~Costa, J.~Penedones, D.~Poland, S.~Rychkov,
[arXiv:1107.3554 [hep-th]].
}

\lref\AnninosUI{
  D.~Anninos, T.~Hartman, A.~Strominger,
[arXiv:1108.5735 [hep-th]].
}

\lref\HaagQH{
  R.~Haag, J.~T.~Lopuszanski, M.~Sohnius,
Nucl.\ Phys.\  {\bf B88}, 257 (1975).
}

\lref\papadodimas{
  S.~El-Showk, K.~Papadodimas,
[arXiv:1101.4163 [hep-th]].
}

\lref\FateevZH{
  V.~A.~Fateev, S.~L.~Lukyanov,
Int.\ J.\ Mod.\ Phys.\  {\bf A3}, 507 (1988).
}
\lref\HeemskerkPN{
  I.~Heemskerk, J.~Penedones, J.~Polchinski and J.~Sully,
  JHEP {\bf 0910}, 079 (2009)
  [arXiv:0907.0151 [hep-th]].
}
\lref\GiombiWH{
  S.~Giombi, X.~Yin,
JHEP {\bf 1009}, 115 (2010).
[arXiv:0912.3462 [hep-th]].
}

\lref\MaldacenaNZ{
  J.~M.~Maldacena, G.~L.~Pimentel,
JHEP {\bf 1109}, 045 (2011).
[arXiv:1104.2846 [hep-th]].
}

\lref\GiombiKC{
  S.~Giombi, S.~Minwalla, S.~Prakash, S.~P.~Trivedi, S.~R.~Wadia, X.~Yin,
[arXiv:1110.4386 [hep-th]].
}

\lref\AharonyJZ{
  O.~Aharony, G.~Gur-Ari, R.~Yacoby,
[arXiv:1110.4382 [hep-th]].
}

\lref\HofmanAR{
  D.~M.~Hofman, J.~Maldacena,
JHEP {\bf 0805}, 012 (2008).
[arXiv:0803.1467 [hep-th]].
}

\lref\BuchelSK{
  A.~Buchel, J.~Escobedo, R.~C.~Myers, M.~F.~Paulos, A.~Sinha and M.~Smolkin,
JHEP\ {\bf 1003}, 111  (2010).
[arXiv:0911.4257 [hep-th]].
}

\lref\DolanDV{
  F.~A.~Dolan and H.~Osborn,
[arXiv:1108.6194 [hep-th]].
}

\lref\VasilievEV{
  M.~A.~Vasiliev,
Phys.\ Lett.\ B\ {\bf 567}, 139  (2003).
[hep-th/0304049].
}

\lref\VasilievBA{
  M.~A.~Vasiliev,
In *Shifman, M.A. (ed.): The many faces of the superworld* 533-610.
[hep-th/9910096].
}

\lref\SezginRT{
  E.~Sezgin and P.~Sundell,
Nucl.\ Phys.\ B\ {\bf 644}, 303  (2002), [Erratum-ibid.\ B\ {\bf 660}, 403  (2003)].
[hep-th/0205131].
}

\lref\BashamIQ{
  C.~L.~Basham, L.~S.~Brown, S.~D.~Ellis and S.~T.~Love,
Phys.\ Rev.\ D\ {\bf 17}, 2298  (1978).
}

\lref\EastwoodSU{
  M.~G.~Eastwood,
Annals Math.\ \ {\bf 161}, 1645  (2005).
[hep-th/0206233].
}

\lref\MikhailovBP{
  A.~Mikhailov,
[hep-th/0201019].
}


\lref\solvayhigher{
Solvay workshop proceedings ,~
http://www.ulb.ac.be/sciences/ptm/pmif/HSGT.htm
}

\lref\BianchiYH{
  M.~Bianchi and V.~Didenko,
[hep-th/0502220].
}

\lref\FranciaBV{
  D.~Francia and C.~M.~Hull,
[hep-th/0501236].
}

\lref\deBuylPS{
  S.~de Buyl and A.~Kleinschmidt,
[hep-th/0410274].
}

\lref\PetkouNU{
  A.~C.~Petkou,
[hep-th/0410116].
}

\lref\BouattaKK{
  N.~Bouatta, G.~Compere and A.~Sagnotti,
[hep-th/0409068].
}

\lref\BekaertVH{
  X.~Bekaert, S.~Cnockaert, C.~Iazeolla and M.~A.~Vasiliev,
[hep-th/0503128].
}

\lref\GiombiVG{
  S.~Giombi and X.~Yin,
JHEP\ {\bf 1104}, 086  (2011).
[arXiv:1004.3736 [hep-th]].
}

\lref\GiombiYA{
  S.~Giombi and X.~Yin,
[arXiv:1105.4011 [hep-th]].
}

\lref\Evans{
N.T.Evans,
J.Math.Phys. 8 (1967) 170-185.}

\lref\KonsteinBI{
  S.~E.~Konstein, M.~A.~Vasiliev and V.~N.~Zaikin,
JHEP\ {\bf 0012}, 018  (2000).
[hep-th/0010239].
}

\lref\KochCY{
  R.~d.~M.~Koch, A.~Jevicki, K.~Jin and J.~P.~Rodrigues,
Phys.\ Rev.\ D\ {\bf 83}, 025006  (2011).
[arXiv:1008.0633 [hep-th]].
}


\hfill
{
PUPT-2399
}

\Title{
\vbox{\baselineskip12pt
}}
{\vbox{\centerline{  Constraining conformal field theories}
\vskip .5cm
\centerline{  with a higher spin symmetry }}}
\bigskip
\centerline{   Juan Maldacena$^a$ and Alexander Zhiboedov$^b$}
\bigskip

\centerline{ \it  $^a$School of Natural Sciences, Institute for
Advanced Study} \centerline{\it Princeton, NJ, USA}

\centerline{\it $^b$Department of Physics, Princeton University}
\centerline{\it Princeton, NJ, USA}

\vskip .3in \noindent

We study the constraints imposed by the existence of a
single higher spin  conserved current  on a three dimensional conformal field theory.
A single   higher spin  conserved  current implies the existence of an infinite number of
higher spin  conserved  currents.
The correlation functions of the stress tensor and the conserved currents are then shown to be
equal to those of a free field
theory. Namely a theory of $N$ free bosons or free fermions.
This is an extension of the Coleman-Mandula theorem to CFT's, which do not have a
conventional S matrix.
We also briefly discuss the case where the higher spin symmetries are ``slightly'' broken.


 \Date{ }



 \listtoc\writetoc
\vskip .5in \noindent

\newsec{Introduction }
\noindent

The classic Coleman-Mandula result \ColemanAD, and its supersymmetric extension
\HaagQH , states that the maximum spacetime symmetry of a theory with an S-matrix
is the super-Poincare group\foot{\HaagQH\ also mentions the conformal group in the case of massless particles. However, \HaagQH\ assumed that these massless particles
are free in the IR, so that an S-matrix exists.  }. Interacting conformal field theories
are interesting theories that do not have an S-matrix obeying the assumptions of
\refs{\ColemanAD,\HaagQH}. In this paper we would like to address the question of
whether a CFT can have a spacetime symmetry beyond the conformal group.
We show that if a CFT has a conserved higher spin current, $s>2$,
 then the   theory is essentially
free. Namely, all the correlators of the conserved currents are those of a free theory.
In particular, this implies that the energy correlation function
 observables \BashamIQ\ of ``conformal collider physics" \HofmanAR\ are
the same as those of a free theory.

Let us clearly state the assumptions and the conclusions.

$\underline{\rm{Assumptions:}}$

\item{a)} The theory is conformal and it obeys all the usual CFT axioms/properties,
such as the operator product expansion, existence of a stress tensor,  cluster decomposition, a finite number of primaries with dimensions less than some number,  etc.
\item{ a')} The two point function of the stress tensor is finite.
\item{b)} The theory is unitary.
\item{c)}  The theory contains a conserved current $j_{s}$ of spin higher than two $s>2$.
\item{d)} We are in three spacetime dimensions.
\item{e)} The theory contains unique conserved current of spin two which is the stress tensor.

$\underline{ \rm{The ~theorem, ~or ~conclusion: }}$

There is an infinite number of even spin
conserved currents that appear in the operator product expansion
of two stress tensor. All correlation function of these
currents have two possible structures. One is identical to that obtained in a theory of
$N$ free bosons, with  currents built as  $O(N)$ invariant bilinears of the free bosons.
  The other is identical to those of a
theory of $N$ free fermions, again with currents given by   $O(N)$ invariant bilinears in the
fermions.

Let us discuss the assumptions and conclusions in more detail.

We spelled out $a')$ explicitly, to rule out theories with an infinite number of degrees
of freedom, as   in  the $N=\infty$ limit of $O(N)$ vector models, for example.
   Unitarity is a very important assumptions since
it allows us to put bounds on the dimensions of operators, etc. We assumed the existence
of a higher spin current.   One might wonder if one could have a symmetry which is not generated
by a local current. Presumably, a continuous symmetry implies the existence of an associated current,
but, to our knowledge, this has not been proven\foot{
Of course, if we have a theory given in terms of a Lagrangian, then Noether's theorem implies the
existence of a current. Also if we assume we can generalize  the action of the symmetry to the
case that the parameter has some spacetime dependence,
$\epsilon(x)$, then the usual argument implies the existence of an associated current, $j_\mu = { \delta
\over \delta \partial_\mu \epsilon } $, where the
derivative is acting on the partition function,
or the generating functional of correlation functions. }.
Assumption $d)$ should hopefully be replaced by $d\geq 3$. Some of the methods in this
paper have a simple extension to higher dimensions, and it should be straightforward
to extend the arguments to all dimensions $d \geq 3$. In two dimensions, $d=2$, we expect
a richer structure. In fact, the Coleman-Mandula theorem in two dimensions allows integrable
theories \ZamolodchikovXM .
Also there are interesting  current  algebras in two dimensions which
contain higher spin primaries beyond the stress tensor. An example are the $W_N$ symmetry
algebras \FateevZH .
The assumption of a unique stress tensor can also be relaxed, at the expense of making the
conclusions a bit more complicated to state.  It is really a technical
assumption that simplifies the analysis.
 In fact, we actually  generalize the discussion
to the case that we have exactly two  spin two conserved currents.
In that case we have a kind of factorization into two subsectors and one, or both, could have
higher spin currents. We expect something similar for a larger number of spin two currents, but
we did not prove it.
A simple theory when there are two spin two conserved currents is the product of a free
theory with a non-trivial interacting theory. In this case all the higher spin currents
live in the free subsector.

Now, regarding the conclusions, note that we did not prove the existence of a free field
operator $\phi$, in the CFT. The reason why we could not do it is easily appreciated by
considering a theory of $N$ scalar fields where we restrict the operators to be $O(N)$
singlets. This is sometimes called ``the free $O(N)$ model''.
This theory obeys all the assumptions of our theorem  (as well as the conclusions!) but it does not have a free field in the spectrum.
Short of establishing the existence of free fields, we will show that the theory
contains ``bilocal'' operators $ B(x_1,x_2)$  and $F_{-}(x_1, x_2)$ whose correlators are the same
as those of the free field operators $ \sum_{i=1}^N :\phi^i(x_1) \phi^i(x_2):$ and $ \sum_{i=1}^N :\psi^i_{-}(x_1) \psi^i_{-}(x_2):$ in a theory
of free fields. Our statements also concern the infinite number of even spin conserved currents
that appear in the operator product expansion of two stress tensors. We are not making any
statement regarding other possible odd spin conserved currents, or spin two currents that
do not appear the operator product of two stress tensors. Such currents, if present,
 are probably also highly
constrained but we leave that to the future.

Notice that we start from the assumption of certain symmetries. By imposing the \Ward identities
 we obtained the explicit form of the correlators. Thus, we can view this as a
simple example of a realization of the bootstrap program. Namely, we never used ``the Lagrangian'',
we derived everything from physical correlation functions and physical consistency conditions.
  Of course, the result turned out
to be rather trivial since we get free theory correlation functions. We can also view this as an exercise in
 current algebra, now for currents of higher spin.

Recently, there have been some studies regarding the duality between various $O(N)$ models
in three dimensions and Vasiliev-type theories \refs{\FradkinQY,\VasilievBA,\SezginRT}
 in $AdS_4$  \refs{\KlebanovJA,\GiombiWH,\GiombiVG} (and $dS_4$ \AnninosUI).
If one  considers a Vasiliev theory  with boundary conditions (at the $AdS_4$ boundary) which
preserve the higher spin symmetry, then the theory obeys the assumptions of our theorem. Thus,
our theorem implies that the theory is equivalent to a free theory of scalars or fermions.
This was conjectured in \KlebanovJA\ (see references therein for previous work), and
 tested in \refs{\GiombiWH,\GiombiVG,\GiombiYA}, see also \refs{\KochCY,\DouglasRC} .
 In this context the quantization of $N$ implies that the coupling constant of a unitary
 Vasiliev's theory is quantized,  if we preserve the higher spin symmetry at the $AdS$ boundary.

It is also interesting to consider the same Vasiliev theory but with boundary conditions
that do not preserve the higher spin charge. An example was proposed in \KlebanovJA\
by imposing a boundary condition for the scalar that produced the ``interacting'' $O(N)$
theory, see also  \GiombiYA . In such $O(N)$ models,
 the higher spin currents acquire  an anomalous dimension of order
$1/N$. Then the conclusions of our theorem do not apply. However, there are still interesting
constraints on the correlators and we discuss some of them, as an example.
It is likely that this method
would give a way to compute correlators in this theory, but we
will leave that for the future.

There is a large literature on higher spin symmetry, and we refer the reader to the
review \solvayhigher , which is also available in the arXiv \refs{\BianchiYH, \FranciaBV, \deBuylPS, \PetkouNU, \BouattaKK, \BekaertVH}.

\subsec{Organization of the paper}

In section two we discuss some generalities about higher spin currents.

In order to make the discussion clear, we will  first present an argument that
rules out theories with operators with low anomalous dimensions. This serves
as a simple warm up example to the arguments presented later in the paper for
the more general case. This is done in section three.

In section four we discuss some general properties of three point functions of currents, which will
be necessary later.

We then  present  two slightly different approaches for showing the main
conclusion. The first, in section five,  requires a slightly more
 elaborate construction but  it is computationally
 more straightforward. The other, in section six,
is  conceptually more straightforward, but it required us to use Mathematica a lot.
We have presented both methods, since  they could be useful for
other purposes (useful spinoffs).
 These two sections can be read almost independently, and the reader should
feel free to choose which one to read or skip  first.

In section seven we discuss the case of slightly broken higher spin symmetry.
We just discuss a couple of simple points, leaving a more general analysis to the future.

In section eight we discuss the case of exactly  two spin two conserved currents.

In section nine we present some conclusions and discussion.

We also included several appendices with some  more technical results, which could also
be useful for other purposes.

\newsec{Generalities about higher spin currents}

We consider higher spin local operators  $J_{\mu_1 \cdots \mu_s}$ that  transform
in the representation of spin $s$ of the rotation group. More specifically, if we consider the operator
inserted at the origin, then it  transforms in the spin $s$ representation. It is convenient to
define the twist,
 $\tau = \Delta - s$, where $\Delta$ is the scaling dimension of the operator. In three dimensions,
the unitarity bound restricts the twist of a primary operator to \Evans
\eqn\twist{\eqalign{
\tau &\geq {1 \over 2}, ~~~~~~{\rm for}~~~ \quad s=0,{1 \over 2} \cr
\tau &\geq 1, ~~~~~~~{\rm for}~~~\quad s \geq 1.
}}

The equality in the first line corresponds to free fields: bosons and fermions.
While in the second line it corresponds to conserved currents $\pa_{\mu} J^{\mu}_{\ \ \mu_2 \cdots \mu_s} = 0$.

From these conserved currents we can build conserved charges by contracting the current of spin $s$
with an $s-1$ index conformal Killing tensor $\zeta^{\mu_1 \cdots \mu_{s-1}}$, which obeys
 $\partial_{(\mu} \zeta_{\mu_1 \cdots \mu_{s-1} )} = 0$%
, where the parenthesis denote the symmetric and {\it traceless} part.
This condition ensures that $\hat J_\mu = J_{\mu \mu_1 \cdots \mu_{s-1} } \zeta^{\mu_1 \cdots \mu_{s-1} } $ is a conserved current so that the associated charge
$ Q = \int_{\Sigma} * \hat J $ is conserved.
A simple way to construct a conformal Killing tensor is to take a product of conformal Killing vectors \EastwoodSU .
In a CFT these charges annihilate the conformal invariant vacuum. These charges are
conserved in the sense that their value does not depend on the hypersurface where we integrate
the current. However, they might not commute with the Hamiltonian, since the Killing vectors
can depend explicitly on the coordinates. This is a familiar phenomenon and it happens already
with the ordinary conformal generators.
These conserved currents lead to identities when they are inserted in correlation functions
of other operators. Namely, imagine we consider an $n$ point function of operators ${\cal O}_i$. Then we will get a ``charge conservation identity'' of the form
\eqn\consei{
 0 = \sum_i \la {\cal O}_1(x_1) \cdots [ Q , {\cal O}_i(x_i)] \cdots {\cal O}_n(x_n) \ra
 }
If we know the action of $Q$ on each operator this leads to certain equations which we call
``charge conservation identities".

In this paper, we will focus exclusively
on only one very particular charge that arises from a Killing
tensor with only minus components. Namely,  only  $\zeta^{- \cdots -  } $ is nonzero, and the rest of the
components are zero. Here $x^-$ is a light cone coordinate.
We think of the three coordinates as\foot{Throughout the paper we consider $x^{\pm}$ as independent variables.} $ds^2 =  dx^+ dx^- + dy^2$.
We denote the charge associated to the spin $s$ current by
\eqn\chargas{
Q_s = \int_{x^+ = {\rm const} }  dx^- dy  ~~ j_s  ~,~~~~~~~~~~ j_s \equiv J_{s,- \cdots - }
}
We have also defined $j_s$ to be the current with all minus indices. Also from now on we will denote by
$\partial \equiv \partial_{x^-} $.

From now on we will consider the twist operator $\tau$ defined so that it is the anomalous dimension
minus the spin in the $\pm $ directions (or boost generator in the $\pm$ direction). Note that $Q_s$ has
  spin $s-1$. One can check that it has twist zero. This is a very important property of this particular
  $Q_s$ charge which we will heavily use.

One more general property that we will need is the following. The fact that $Q_s$ has a non-trivial
dimension $(s-1)$ under the dilatation operator implies that
\eqn\qss{
[ Q_s , j_2 ] \propto \partial j_s + \cdots  ~,~~~~~~~~~~~~~ s>1
}
(recall that $\partial = \partial_-$). This is indicating that the current $j_s$, from which we formed the
charge $Q_s$ is appearing in the right hand side of the commutator \qss .
 Here $j_2$ is the stress tensor of the theory.
This is shown in more detail in appendix  A. This fact is also related to the following. For any operator
${\cal O}$, the three point function $\langle {\cal O} {\cal O} j_2 \rangle$ is nonzero, where $j_2$ is
the stress tensor. This should be non-zero because the stress tensor generates   conformal transformations.
In particular, the stress tensor should always be present in the OPE of two identical operators. Here
we are using that the two point function of the stress tensor is finite. (The stress tensor comes with a
natural normalization so that it makes sense to talk about the coefficient in its two point function).

Note that two point functions of conserved currents are proportional to
\eqn\twopo{
\langle j_s(x_1) j_s(x_2) \rangle  \propto {  (x^+_{12})^{2 s } \over |x_{12}|^{ 4 s + 2} }
}
(Recall that  our definition of $j_s$ \chargas\ contains only the minus components.)

\newsec{Removing operators in the twist gap, $ { 1 \over 2 } < \tau < 1$}

Notice that  all operators with spin $s \geq 1 $ should have twist bigger or equal to  one. However, operators with
spin $s=0,{1\over 2}$ can have   twists less than one.   If $\tau =1/2$, we
 have a free field which can be factored out of the theory. In this section we show that it is very
easy to eliminate operators with twists in the range
 $ {1 \over 2 }< \tau <1$. We call this range the ``twist gap''.
  This will serve as a warm up exercise for the rest of the paper.

To keep the discussion simple, let us assume that we have
 a  conserved current of spin four, $j_4$. (We will show in the next section that a current of spin four is
 always present as soon as we have {\it any} higher spin current).
Let us see how this current could act on a scalar operator of spin zero $\phi $ and twist in the twist gap,  $ { 1 \over 2 } < \tau < 1$.
The action of the charge $Q_4$, $[Q_4 , \phi ] $,   preserves the twist. $Q_4$ cannot annihilate $\phi $ since
the operator product expansion of $\phi \phi  $ contains the stress tensor (the unique
spin two conserved current), and $Q_4$ cannot
 annihilate the stress tensor due to \qss . The right hand
side of $[Q_4 , \phi ] $ should be a combination of local operators and derivatives. We cannot have explicit functions of
the coordinates since that would imply that $Q_4$ does not commute with translations. The fact that
$Q_4$ commutes with translations can be seen from  its integral expression \chargas\ and the
conservation of the current.
The operator that appears in the right
hand side has to have the same twist as $\phi $. Thus, it should be a scalar. The only derivative that
does not change the twist is $\partial = \partial_-$. This derivative should be present three times due
to spin conservation.
In conclusion, we have that
\eqn\actq{
[Q_4 , \phi  ]  = \partial_-^3 \phi
}
If we had many scalar operators of the same twist we would get $[Q_4,\phi _a] = c_{ab} \partial^3 \phi _b$,
where $c_{ab}$ is symmetric due to the \Ward identities for  $Q_4$ acting on $\langle\phi _a \phi _b \rangle \propto \delta_{ab}$.
Thus, we can diagonalize $c_{ab}$ and we return to \actq .

As the next step we can write the \Ward identity for the action of $Q_4$ on the
four point functions of these fields. It is convenient to do it in momentum space. Then we have
\eqn\WIms{\eqalign{
[\sum_{i=1}^{4}(k_{i-})^3 ]\la \phi(k_1) \phi(k_2) \phi(k_3) \phi(k_4) \ra &= 0
}}
Using  momentum conservation, and writing $k_{i} = k_{i-}$ for the minus component of the momentum, we find
\eqn\momcons{
\sum_{i=1}^{4} k_{i}^3 = - 3 (k_1 + k_2) (k_1 + k_3) (k_2 + k_3)   = 0.
}
when it acts on the four point function.
Thus, the four point function is  proportional to  $\delta(k_{i_1} + k_{i_2}) \delta(k_{i_3} + k_{i_4})$ and other two
terms obtained by permutations. Together with rotational invariance and conformal invariance,
this means that the four point function is a sum of two point functions
\eqn\solution{\eqalign{
\la  \phi (x_1) \phi(x_2) \phi(x_3) \phi(x_4) \ra &=\la \phi(x_1) \phi(x_2) \ra \la \phi(x_3) \phi(x_4) \ra + {\rm permutations}
}}
Now we will show that the only operators that can obey
\solution\ are free fields (see \papadodimas ). For that purpose we perform the ordinary operator product expansion as $|x_{12}| \to 0$.
The first term in \solution\ gives  the contribution of the identity operator. The other two terms
 are analytic around $x_{12} =0$. Thus, they correspond to operators with twists  $\tau_{int} =
2 \tau  + n $, with $n\geq 0$, where $\tau$ is the twist of $\phi$. If this twist is $\tau>1/2$, then the
stress tensor, which has twist one,
 would not appear in this operator product expansion. But the stress tensor should always
 appear in the OPE of two identical currents.
Thus, the only possible value is $\tau =1/2$ and we have a free field. Such a free field will decouple from
the rest of the theory.

Actually, by a similar argument we can eliminate spin $1/2$ operators in the twist gap,
$ { 1 \over 2 } < \tau < 1$. We repeat the above
arguments for the particular component $\psi = \psi_{-}$ (the spin 1/2 component as opposed to the spin -1/2).
All the arguments go through with no change up to \momcons . That equation restricts the $x^-$ dependence
only. However, conformal symmetry and rotational invariance allow us to compute the four point function
for any component $\psi_{\alpha}$, and we find an answer that factorizes. Which then implies again $\tau =1/2$
and a free fermion.

Note also that the analysis leading to \actq\ also constrains how all higher spin charges act on free fields,
$[Q_s , \phi ] = \partial^{s-1} \phi $, and similarly for fermions (we can show it easily for the $\psi_-$
component and then the Dirac equation ensures that
it acts in the same way on both components $\psi_{\alpha}$\foot{Here we again use that $[Q_{s}, P_{\mu}] = 0$.}).

At this point we should mention a qualitative argument for the reason that we get free fields. If
we form wavepackets for the fields $\phi$  centered around some momentum and somewhat localized
in space, then the action of the charge $Q_4$ will displace them by an amount which is proportional
to the square of their momenta. Thus, if the wavepackets were colliding in some region, then these
displacements would make them miss. Here we use that  $d\geq 3$.

The results of this section can be very simply extended to $d\geq 3$. There the twist gap exists in
the region ${ d- 2 \over 2 } \leq \tau < d -2$, with the lower bound corresponding to free fields.

Returning to $d=3$,
in what follows, we will mostly consider the twist one operators that correspond to conserved currents.
The idea will be similar, we first constrain the action of the higher spin charges and
then determine the correlation functions of the operators.

We will present two independent ways of doing this. The first involves the notion of light cone operator
product expansions and it is a bit more direct. It is presented in section five.  The second  involves
a more explicit analysis of three point functions and it is conceptually easier, but computationally
more complicated (needs use of Mathematica). It is presented in section six.
 The reader can choose which one to read and/or skip.
But first some further generalities.

\subsec{Action of the charges on twist one fields}

Before jumping to those sections, let us make some preliminary statements on the action of charges $Q_s$
on twist one operators. Since $Q_s$ has twist zero, the only things we can have in the right hand side
are other twist one fields. These include other (or the same) twist one fields. Even if we had fields in
the twist gap (${ 1 \over 2 } < \tau < 1$), they cannot
appear in the right hand side of a twist one current transformation, due to
twist conservation and the fact that derivatives can change the twist, at most by integer amounts.
Thus, we have the general transformation law
\eqn\gen{
[Q_s , j_{s'} ] = \sum_{s'' = {\rm max}[s'-s+1,0]}^{s'+s-1}  \alpha_{s,s',s''} \  \partial^{s' + s -1 - s''} j_{s''}
}
(the limits of the sum are explained below).  Here we wrote explicitly arbitrary constants $\alpha_{s,s',s''}$ appearing in front of each term, below we will assume them implicitly.  We will also sometimes omit the
derivatives.
The sum \gen~ can involve twist one and $s=0,1/2$ fields, which are not conserved currents, but we still denote them
by $j_0, ~j_{1/2}$. Here everything has minus indices which we have not indicated. The derivatives are also minus derivatives. The number of derivatives is easily fixed by matching the spin on both sides.

One easy property we can prove is the following, if current $j_y$ appears in the right hand side of $[ Q_s , j_x]$,
then we should have that current $j_x$ appears in the right hand side of $[Q_s , j_y]$. This follows by considering
the action of $Q_s$ on  the $\langle j_x j_y \rangle$ two point function and \twopo .
 In summary,
 \eqn\jxjy{
 [ Q_s , j_x]  = \partial^{s+ x - y -1} j_y + \cdots ~~ \Rightarrow ~~~[Q_s , j_y] = \partial^{s+y- x -1} j_x + \cdots
 }
 This has a simple consequence
which is the fact that the spread of spins $s''$ in \gen\ is as indicated in \gen . The upper limit in \gen\
 is
obvious. The lower limit in \gen\ arises from the fact that the current $j_{s'}$ should appear in the right hand side of
$[Q_s , j_{s''}]$, and the upper limit in this commutator, results in the lower limit in \gen .

\newsec{Basic facts about three point functions}

For our analysis it is important  to understand the structure of the three point functions
of conserved currents and other operators in $d=3$.

This problem was analyzed in \GiombiRZ\ (see also \CostaMG ).
Unfortunately, it is not clear to us what was proven and what was the result of
a case by case analysis with a later extrapolation (a ``physics'' proof).
Thus, in appendix I we prove the
results that will be crucial for section five. These only involve certain light-like limits of the correlators.
For section six we need some particular cases with low spin, where the statements in
\refs{\GiombiRZ,\CostaMG} can be explicitly checked.

Nevertheless, let us summarize the results of \GiombiRZ .
The three point function of the conserved currents  has, generically,  three
distinct pieces
\eqn\threegen{
\la j_{s_1} j_{s_2} j_{s_3} \ra = \la j_{s_1} j_{s_2} j_{s_3} \ra_{boson} +
\la j_{s_1} j_{s_2} j_{s_3} \ra_{fermion} + \la j_{s_1} j_{s_2} j_{s_3} \ra_{odd}.
}

Here the piece $\la j_{s_1} j_{s_2} j_{s_3} \ra_{boson}$ is generated by the theory of free bosons. The piece $ \la j_{s_1} j_{s_2} j_{s_3} \ra_{fermion}$ is generated by the theory of free fermion and the piece $ \la j_{s_1} j_{s_2} j_{s_3} \ra_{odd}$ is
not generated by   free theories. The boson and fermion pieces are known in closed form while the odd piece can be
computed in each particular case explicitly by imposing the conservation of the currents.
By considering particular examples it was observed that the odd piece is non-zero whenever the triangle rule is satisfied
\eqn\trianglerule{
s_{i} \leq s_{i+1} + s_{i+2}.
}
In appendix B  we derive an integral expression for these odd correlators that naturally
incorporates this triangle rule. We start with free higher spin
conformal fields in four dimensions, introduce a conformal invariant perturbation,  and
look at correlators on a three dimensional slice. The correlators automatically obey the
conservation condition and are conformal invariant. See appendix B  for more details.

Another property of three point functions which we found useful is  that any three point function of the two identical\foot{We need literally the same current. It is not enough that they have
the same spin.} currents with
the third one that has an odd spin is zero
\eqn\basicfact{
\la j_s j_s j_{s'} \ra = 0 ~~~~~~~~~~~{\rm for }~~s' ~~{\rm odd}
}
 This is easy to check for the explicit expressions for the boson and fermion
solutions. It is also can be seen from the integral representation for the odd piece.


\subsec{General expression for three point functions  }

The space of three point functions of conserved currents is conveniently split into
two parts according to their transformation under the parity. Practically, we call ``even'' ones the three point functions
that do not contain the three dimensional epsilon  tensor and ``odd'' the ones that do.

The parity  even correlation functions are generated by
\eqn\genfunc{
{\cal F}_{even}  = e^{{1 \over 2} (Q_1 + Q_2 + Q_3)} e^{P_1 + P_2} (b \cosh P_3 + f \sinh P_3 )
}
where the piece proportional to  $b$ stands for correlation functions in the theory of free bosons and
the one proportional to  $f$ for the ones in the free fermion theory\foot{An equivalent, but  symmetric version of \genfunc\ is given by $${\cal F}_{even}  = e^{{1 \over 2} (Q_1 + Q_2 + Q_3)} e^{(P_1 + P_2+P_3)} \times
\left(b+f + {b - f \over 3} [e^{- 2 P_1} + e^{- 2 P_2} + e^{- 2 P_3}] \right) $$}.
The $P_i$ and $Q_i$ are some cross ratios whose form can be found in   \GiombiRZ .
For currents which have indices only along the minus direction we can find the
simple expressions
\eqn\simplr{
Q_i = \ell_i^2 (   \hat x^+_{i,i+1} - \hat x^+_{i,i-1} ) ~,~~~~ P_i = \ell_{i+1} \ell_{i-1} \hat x^+_{i-1,i+1} ~,~~~~~~ \hat x^+ = { x^+ \over x^2 }
}
where $\ell_i$ are the parameters in the generating function. In other words the term with $\ell_1^{2 s_1}$ gives us the current $j_{s_1}(x_1) $, etc.

When spins of the currents are integer \genfunc\ reproduces the results given in  \GiombiRZ , but
\genfunc\ also gives the answer for half integer spin currents\foot{Note that for half integer
spins the labels $b$ and $f$ just give two different possible structures, since, of course
we need both bosons and fermions to have half integer spins}.

Due to the different behavior under $P_{3} \to - P_{3}$ the
bosonic and fermionic parts never mix inside the
\Ward identities for three point functions that we will consider.
 Thus, whenever we  can solve  the \Ward identities they are satisfied
  separately for $f=0$ and $b=0$ parts.

\newsec{Argument using bilocal operators }

Here we present an argument for the main conclusion.

\subsec{Light cone limits of correlators of conserved currents}

We consider the light cone OPE limit of two conserved currents $j_s(x) j_{s'}(0)$.
All indices are minus.
We will consider the twist one contribution to the OPE\foot{Here we consider
the contribution of twist {\it exactly} one. This should not be confused with discussions in
weakly coupled theories where one is interested in the whole tower of operators which have
twists close to one (or in four dimensions twist close to two). Such operators appear
in discussions of deep inelastic scattering and parton distribution functions. }. This can be cleanly separated from the lower
twist contribution which can only arise from the identity or spin $s=0, {1\over 2 }$ operators.
This clean separation is possible because there is a finite number of operators in the twist
gap, ${ 1 \over 2} < \tau < 1$.
To extract the twist one contribution of a given spin,
 we can consider the three point functions
  $\langle j_s(x_1) j_{s'}(x_2) j_{s''} (x_3) \rangle$. We can also consider the contribution
  of all spins together by taking the following light-like limits.

We take the limit $x^+_{12} \to 0$ first.
Then we take the limit $y_{12}\to 0$. By twist conservation,
the correlator could behave only like $1/|y_{12}|$ or like $1/y_{12}$, in this limit.
The fermion piece vanishes when we take $x_{12}^+\to 0$.
One can see that the $1/y_{12}$ behavior can be produced only by a parity odd piece. This follows
from the fact that we can put the third current at $y_3 =0$, but at generic $x^\pm_3$. This still allows
us to extract any possible twist one current that appears in the OPE limit. However, the parity
transformation $y \to -y$ would change the sign of the correlator only if we have a $1/y_{12}$ term
in the OPE.
Below, the part that contains the  $1/|y_{12}|$ behavior is called the ``boson'' piece. It comes from
the boson piece of the three point functions.

Similarly, we can look at the piece that goes like $x_{12}^+$ in the limit that $x_{12}^+ $ goes to
zero. Then we extract the piece going like ${ x_{12}^+ \over |y_{12}|^3 }$.

More explicitly, these limits are defined by
\eqn\limitsdef{ \eqalign{
\underline{ j_s j_{s'}}_b = &  \left( \lim_{
 y_{12} \to 0^+ }  + \lim_{y_{12} \to 0^-} \right) |y_{12}| \lim_{x_{12}^+ \to 0 } j_s(x_1) j_{s'}(x_2) - ({\rm lower})
 \cr
  \underline{ j_s j_{s'}}_f = &   \left(  \lim_{
 y_{12} \to 0^+ }  + \lim_{y_{12} \to 0^-} \right)  \lim_{x_{12}^+ \to 0 } { |y_{12}|^3 \over x_{12}^+}
 \left[  j_s(x_1) j_{s'}(x_2) - { 1 \over |y_{12}| } \underline{ j_s j_{s'}}_b  - ({\rm lower}) \right]
 }}
 where we have indicated by an underline the fact that we take the light-like limit (and
 the subindex on the line reminds us which of the two limits we took). We will also indicate
 by an underline points that are light-like separated along $x^-$.
 In the fermion like limit \limitsdef ,  we have indicated that we extracted
 the boson piece. However, in practice we will
 only use it when the boson piece  vanishes. Thus, there is no ambiguity in extracting
 the limit. We also extract any possible twist less than one operators that could appear. There is a finite list of them. Thus, there
is no problem in extracting the lower twist operators.
In section three we have eliminated these lower twist operators by making the assumption
of the existence of $Q_4$. Here we have not yet derived the existence of $Q_4$, so we had
to argue that these lower twist operators do not affect the definition of the limits and the
extraction of the twist one contribution to the OPE. In fact, from now on, these lower twist
contribution play no role. After we show the existence of $Q_4$, we can then remove them using
section three.
 We could define also a limit that extracts the odd piece, but we will not need it.

The three point functions simplify dramatically in this limit.  For the boson part
we have
\eqn\limitbcorr{\eqalign{
\left(  \lim_{
 y_{12} \to 0^+ }  + \lim_{y_{12} \to 0^-} \right) \lim_{x^{+}_{12} \to 0} | x_{12}| \la j_{s_1} j_{s_2} j_{s_3} \ra_{b} &  \propto \pa_{1}^{s_1} \pa_{2}^{s_2} \la \underline {\phi \phi^*}  j_{s_3} \ra_{free} ~;
\cr
\la \underline {\phi \phi^* } j_{s_3} \ra_{free} \propto
& {1 \over \sqrt{ \hat{x}_{13} \hat{x}_{23} } } \left({ { 1 \over \hat{x}_{13} } - { 1 \over  \hat{x}_{23} } }\right)^{s_3}
}}
where $\langle \underline {\phi \phi^* } j_{s_3} \rangle_{free}$ stands for the correlation function
of a  free complex boson. The underline reminds us that the points are light-like separated, but there is no limit involved.
The reason we take a {\it  complex} boson is to allow for non-zero values when
 $s$ is odd.
 We have displayed only the $x_i^-$ dependence\foot{There is also a factor of ${ 1 \over x_{13}^+ } = { 1 \over x_{23}^+ } $ in the right hand side of \limitbcorr . }  and we have defined a
slightly shifted version of the coordinates
\eqn\hatx{\eqalign{
\hat{x}_{1}  &= x_{1 }^- ~,~~~~~~~ \hat{x}_{2}  = x_{2 }^- ~,~~~~~~~~~~~~\hat x_3 = x_3^- - {y_{13}^2 \over x^{+}_{13}}
}}
We should emphasize that in \limitbcorr\ the expression $\la \phi \phi^* j_{s_3} \ra_{free}$ denotes
 a correlator in a theory of a free complex boson. In particular, $j_{s_3}$ is not the
 original current in the unknown theory, but the current of the free boson theory.

For a free boson theory,  it is easy to see why \limitbcorr\  is true. For a free boson the currents are
given by expressions like $\sum \partial^i \phi \partial^{s-i } \phi^*$. When we
take the limit $x^+_{12} \to 0$ only one term survives from the two sums associated to $j_{s_1}$
and $j_{s_2}$ (recall that $\partial = \partial_-$). The reason is that we need a contraction of the scalar with no derivatives in
order {\it {not} } to bring down a factor $x_{12}^+$ in the numerator. That single term
contains all the derivatives on the $\phi$ fields that are contracted with $j_{s_3}$.

Actually, without assuming the explicit forms given in \GiombiRZ , one can prove that
\limitbcorr\ follows, in the light cone limit, from conformal symmetry and current conservation
for the third current. Since this is a crucial point for our arguments, we provide an explicit proof of \limitbcorr\
in appendix I.

Using the same reasoning for the fermion part we obtain (see appendix I)
\eqn\limitfcorr{\eqalign{
\left(  \lim_{
 y_{12} \to 0^+ }  + \lim_{y_{12} \to 0^-} \right) \lim_{x^{+}_{12} \to 0} {  |x_{12}|^3 \over  x_{12}^{+} } \la j_{s_1} j_{s_2} j_{s_3} \ra_{f}  & \propto \pa_{1}^{s_1 - 1} \pa_{2}^{s_2 - 1} \la \underline { \psi \psi^* } j_{s_3} \ra_{free}  ~;
 \cr
\la \underline {\psi \psi^* } j_{s_3} \ra_{free}   & \propto      {1 \over ( \hat{x}_{13} \hat{x}_{23} )^{3/2} }  \left({ { 1 \over \hat{x}_{13} } - { 1 \over  \hat{x}_{23} } }\right)^{s_3-1}
}}
where $\psi$ stands for a free complex  fermion. If $s=0$, then  $\la \underline { \psi \psi^* } j_0 \ra_{free}=0$, and the limit in the first line of \limitfcorr\ vanishes. This is simply the statement that
in a free fermion theory we do not have a twist one spin zero operator.

As an aside, for the odd piece one can obtain a similar expression, given in appendix B.
This will not be necessary here, since the higher spin symmetry will eliminate the odd piece and
we will not need to know its explicit form.

\subsec{Getting an infinite number of currents}

Imagine we have a current of spin $s$. From \qss\  we know that
  $[Q_s , j_2] = \partial j_s  + \cdots $. Then \jxjy\ implies that
    \eqn\jtwo{
    [Q_s , j_s] = \partial^{2 s -3} j_2 + \cdots
    }
    Let us first
    \eqn\assump{
    {\rm assume ~ that ~~} \la j_2 j_2 j_2 \ra|_b \not =0
    }
We now consider the \Ward identity for $Q_s$ acting on
$ \langle \underline { j_2 j_2}_b  j_s \rangle$, where  we mean a \Ward identity that
results from acting with $Q_s$ on $\la j_2 j_2 j_s \ra$ and then taking the light cone
limit for the first two variables. In other words, we get
\eqn\reswa{ \eqalign{
& 0=\la \underline {  [Q_s , j_2]  j_2 }
  j_{s} \ra + \la \underline { j_2  [Q_s , j_2]  }  j_{s} \ra +\la \underline {   j_2  j_2 }  [Q_s ,j_{s} ]\ra
}}
In the first two terms we first expand the commutator using \gen\ and then take the
light cone limit using \limitbcorr .
We will consider first  the case of integer $s$.
Then the  first two terms in \reswa , produce
 \eqn\expresf{
 \la \underline {  [Q_s , j_2]  j_2 }
  j_{s} \ra + \la \underline { j_2  [Q_s , j_2]  }  j_{s} \ra = \partial_1^2 \partial_2^2
 \left[  \gamma  \partial_1^{s-1} + \delta \partial_2^{s-1} \right] \langle \underline { \phi \phi^* } \,j_s \rangle_{free} }
 Here the symmetry under $x_1 \leftrightarrow x_2$ implies that $\gamma = (-1)^s \delta $,
 due to the symmetries of \limitbcorr .
In the third term in \reswa ,  when $Q_s$ acts on $j_s$, we  generate many
 possible operators through \gen . Each of those
 currents  might or might not have an overlap with $\underline { j_2 j_2 }_b$. Thus, combining this with \expresf\ we  conclude that the \Ward identity that results from
 acting with $Q_s$ on $ \langle  \underline { j_2 j_2 }_b    j_s \ra $ is
 \eqn\wards{
 0 = \partial_1^2 \partial_2^2 \left[ \gamma (  \partial_1^{s-1} + (-1)^s  \partial_2^{s-1}  ) \langle
\underline {\phi \phi^* } \,  j_s \rangle_{free} + \sum_{k =1}^{ 2 s -1}  \tilde \alpha_k   \partial_3^{ 2 s -1 - k }
 \langle  \underline { \phi \phi^*  }  \,  j_k \rangle_{free} \right]
 }
 where $\tilde \alpha_k$ is the product of the constants in \gen\ and the actual value of the boson
 part of the three point function. The overall derivatives can be removed since the right hand
 side is a sum of terms of the form $ { 1 \over x_{13}^{a} x_{23}^b }$ where $a$ and $b$ are
 half integer, so if a term was non-zero before the derivative acted it would be non-zero afterwards,
and terms with different powers cannot cancel each other.
Notice that \wards\  involves the explicit functions defined in \limitbcorr . An important property
of these functions is that  $\la \underline { \phi \phi^*}  j_k \ra_{free} $ has a zero of order $k$ when
$x^-_{12} \to 0 $, and the $\partial_3$ derivative does not change the order of this zero.
 This implies that all the terms in the last sum in \wards\ are independent. Thus,
 all the $\tilde \alpha_k$ are fixed.
  We know that $\tilde \alpha_2$ is nonzero due to \jtwo\ and
 \assump . Here we used that the spin two current
 is unique so that there is no other contribution that could
cancel the $j_2$ contribution.
 Given that this term is non-zero, then $\gamma $ is nonzero.  The $\tilde \alpha_k$ for odd
$k$ are set to zero by the $x_{1} \leftrightarrow x_2$ symmetry of the whole original equation,
\reswa . The rest of the $\tilde \alpha_k$ are fixed, and are equal to what we would obtain in a free theory,
which is the unique solution of \wards . It is then possible to check
that  all  $\tilde \alpha_k$,  with
  $k= 2, 4, \cdots ,2 s -2$  are non-zero.  The explicit proof of this statement is given in appendix J.

   In particular,  this implies that the current $j_4$ is in the
spectrum, and it is in the right hand side of the $\underline{j_2 j_2}_b$ OPE.  This is true regardless of
whether the original $s$ is even or odd. Now we can go back to section three and remove the
operators in the twist gap, ${ 1 \over 2} < \tau < 1$.

Since $s>2$, we find that $2 s -2 > s$,
so that we are also finding currents with spins bigger than the original one. Thus, repeating
the argument we find an infinite number of conserved currents.

 We now  show that   a twist one  scalar $j_0$
is present in the spectrum,  and in the right hand side of $\underline {j_2 j_2}_b$. This can be shown by
considering the $Q_4$ \Ward identity acting on $\langle \underline {j_2 j_2 }_b j_2 \ra$.
This leads to an expression very similar to \wards , but with different limits in the
sum (see \gen ), such that now a non-zero $\tilde \alpha_4$ implies the existence of $j_0$.
Now, considering the
ward identity from $Q_4$ on $\langle \underline{j_2 j_2}_b j_0 \rangle$ we show that
$j_2$ is in the right hand side of
\eqn\qfourz{
 [Q_4, j_0 ] = \partial^3 j_0 + \partial j_2 + \cdots
}
 where
the dots denote  other twist one operators  that have no overlap
with $ \underline { j_2 j_2 }_b $ which could possibly appear.
Let us now show that the fermion components are zero.
We  consider the \Ward identity from $Q_4$ on $\langle \underline{j_2 j_2}_f j_0 \rangle $.
Notice that $\langle \underline { j_2 j_2 }_f  j_0 \rangle  =0$, see \limitfcorr .
 Thus, this \Ward identity implies that $\langle j_2 j_2 j_2  \rangle_f =0$. Here we have used that there is a unique stress tensor since we assumed that
the $j_2$ appearing at various places is always the same.
Thus, if $\langle   j_2 j_2 j_2 \rangle_b \not=0$, then
$\langle j_2 j_2 j_2  \rangle_f =0$. Conversely,
 if $\langle j_2 j_2 j_2  \rangle_f \not =0$, then
 $\langle  j_2 j_2 j_2 \rangle_b =0$,  and the discussion after  \wards\ implies that
 also $\la \underline { j_2 j_2 }_b j_s \ra=0$ for all $s$.

In the case that $\la j_2 j_2 j_2 \ra_f$ is nonzero, we can
start with a ward identity similar to \wards , but for
$\langle \underline{j_2 j_2}_f j_s \rangle$. We find an identical conclusion, an infinite set of even spin
currents in the right hand side of $\underline { j_2 j_2 }_f$. (See appendix J for an explicit
demonstration).
In conclusion, due to the uniqueness of the stress tensor we have only one of two cases:
\eqn\casest{\eqalign{
\langle j_2 j_2 j_2 \rangle_b \not = & 0 ~,~~\Rightarrow \langle j_2 j_2 j_2 \rangle_f   =0 ~,~~ \underline { j_2 j_2 }_b   = \sum_{k=0}^\infty[  j_{2 k} ] ~,~~\underline { j_2 j_2 }_f   = 0
\cr
 \langle j_2 j_2 j_2 \rangle_f \not = & 0 ~,~~\Rightarrow \langle j_2 j_2 j_2 \rangle_b   =0 ~,~~ \underline { j_2 j_2 }_f   = \sum_{k=1}^\infty[  j_{2 k}]  ~,~~\underline { j_2 j_2 }_b   = 0
}}
where the brackets denote currents and their derivatives. In other words, the brackets are the
contribution of the conformal block of the current $j_s$ to the twist one part of  the OPE.

It is also possible to start with a half integer $s$, higher spin conserved
 current $j_s$ and to obtain
a \Ward identity similar to \wards \ (we describe it in  detail in appendix D).
 This again shows that   currents with even spins $k = 2, 4 , \cdots 2 s -1$ appear in the
 right hand side of $\underline { j_2 j_2  }_b$ or $\underline { j_2 j_2 }_f $. Again,  if
 $s\geq 5/2$ we get higher spin  conserved currents with even spins
 and we return to the previous case.\foot{
If we have a half integer higher spin current, we will also have
a supercurrent, a spin 3/2 current. This then requires that both
$\la j_2 j_s j_2 \ra_b $ and $\la j_2 j_2 j_2 \ra_f$ are nonzero.
 Since we had the dichotomy \casest ,  we should have more than one spin two conserved
 current if there is
any  half integer higher  spin current.  }

In the next two subsections we consider the first case in \casest\ and then the second.

\subsec{Definition of bilocal operators}

In the  light-like limit we encountered a correlator which was essentially given
by a product of free fields \limitbcorr . Here we will argue that we can
``integrate'' the derivatives and define a bilocal operator $B(x_1,x_2)$ with
$x_1$ and $x_2$ separated only along the minus direction.

Now, let us make a comment on light cone OPE's. The light cone OPE of two
currents $\underline { j_s j_{s'}}_b$, contains only twist one fields of various spins.
The overlap with each individual current is computed by \limitbcorr . In particular,
all such limits will contain a common factor of $\partial_1^{s} \partial_2^{s'} $.
We can ``integrate'' such derivatives and define a quasi-bilocal operator $\hat B ( \underline {x_1 ,x_2} )$
such that
\eqn\bilocsp{
\underline { j_s j_{s'}}_b = \partial_1^{s} \partial_2^{s'} \hat B( \underline {x_1 ,x_2}  )
}
The underline reminds us that  $x_1$ and $x_2$   are null
separated (along the $x^-$ direction).
Here the right hand side is simply a superposition of twist one fields and their derivatives.
They are defined such that
\eqn\formlc{
 \langle \underline { j_s j_{s'}}_b j_{s''}(x_3) \rangle = \partial_1^{s} \partial_2^{s'}
 \langle \hat B(\underline {x_1 ,x_2} ) j_{s''}(x_3) \rangle
 }
 Of course, we also see that $\la \hat B(\underline {x_1 ,x_2}  ) j_{s}(x_3) \ra \propto \langle
\underline { \phi(x_1)
 \phi^*(x_2)  }  j_{s_3}(x_3) \ra_{free} $.
  This implies that $\hat B$ transforms as two weight $1/2$ fields
 under conformal transformations.

A particular one we will focus on is the quasi-bilocal that we get from the stress tensor,
defined by
\eqn\correc{
 \underline{j_2j_2}_b = \partial_1^2 \partial_2^2 B(\underline{x_1,x_2})
 }
These are reminiscent of the (Fourier transform along $x^-$ of the)
operators whose matrix elements define parton distribution functions. One very important
difference is that here we are constructing $B$  {\it only} from the operators that have
twist {\it exactly} equal to one.

We can define similar operators in the case of fermions
\eqn\correcf{\eqalign{
\underline { j_s j_{s'}}_f &= \partial_1^{s-1} \partial_2^{s'-1} \hat F_{-}( \underline {x_1 ,x_2}  ), \cr
\underline{j_2j_2}_f &= \partial_1 \partial_2 F_-(\underline{x_1,x_2})
}}
 Here $F_-$ transforms as the product of two free fermions
with minus polarizations $:\psi_{-}(x_1 )\psi_{-}(x_2):$. It is defined again through
 \limitfcorr , by ``integrating'' the derivatives. It is the superposition of all the
 (even spin) currents that appear in the right hand side of $\underline {j_2j_2}_f $.

We call these operators ``quasi-bilocals''  because they transform under conformal
transformations as a product of two elementary fields. This does not mean that they
are honest local operators, in the sense that their correlators have the properties of
products of fields, with singularities only at the insertion of other operators.
This is illustrated in appendix H.
 Our task will be to show that in a theory with higher spin symmetry, they become
 true bilocal operators, with correlators equal to the free field ones.
 The first step is to constrain the action of $Q_s$ on $B$ and $F_-$ in
  \correc , \correcf .

  \subsec{Constraining the action of the higher spin charges}

 Here we will show that
 \eqn\propeqn{
  [ Q_{s} , B( \underline{x_1,x_2})] = ( \partial_1^{s-1} + \partial_2^{s-1} ) B(\underline{x_1,x_2})
  }
 Then we will show that this implies that their correlators have the free field form.
Let us first assume that  $\langle j_2 j_2 j_2\rangle_b$ is nonzero.
 In \propeqn\ we consider charges $Q_s$ with $s$ even constructed out of the currents $j_s$
that appear in the right hand side of
   $\underline {j_2 j_2}_b$. As we saw before, there is an infinite
  number of such currents.

We would like to compute $[Q_s, B(\underline{x_1,x_2})]$.
We can compute $[Q_{s} , \underline{j_2,j_2}_b] =
\underline{ [Q,j_2] j_2 }_b + \underline{j_2 [Q,j_2] }_b $. In other words, the action of
$Q_s$ commutes with the limit.
This follows from the \Ward identity of $Q_s$ on $\la j_2 j_2 j_k \ra $ and taking the light cone limit, which
gives
\eqn\commutw{
 \la \underline { [Q_s , j_2 ] j_2 } j_k \ra + \la  \underline { j_2 [Q_s , j_2 ] } j_k \ra = - \la
\underline { j_2 j_2 } [Q_s , j_k ] \ra = \la [Q_s , \underline { j_2 j_2 } ] j_k \ra
}
 So,  we can write $[Q,j_2] $ in terms of currents and derivatives
(with indices and derivatives all along the minus directions). Thus, in the end we
can use the formula \formlc\ to write
  \eqn\comput{
   [ Q_s, B(\underline{x_1,x_2}) ] = ( \partial_1^{s-1} + \partial_2^{s-1}) \tilde B(\underline{x_1,x_2}) + ( \partial_1^{s-1} - \partial_2^{s-1})   B'(\underline{x_1,x_2})
   }
   $\tilde B$ contains all the even
   currents and it is symmetric under the interchange of $x_1$ and $x_2$ and $B'$ contains the
   odd currents and it is antisymmetric under  the interchange.  Of course, the full
   expression is symmetric under this interchange.

   Let us first show that $B'$ is zero. Since $B'$ is odd,
   it includes only odd currents $j_{s'}$ with $s'$ odd.
     Imagine that $B'$ contains a particular odd current $j_{s'}$. Then, if $s'>1$, we
     consider the \Ward identity coming from the action of $Q_{s'}$ on $\langle B' j_2 \rangle $
     \eqn\wardint{ \eqalign{
     0 = & \la [Q_{s'}, B' ] j_2 \ra + \la  B' [ Q_{s'},j_2] \ra
     \cr
     0=&  \gamma ( \partial^{s'-1} - \partial^{s'-1} ) \la \underline {\phi \phi^*} j_2 \ra  + \sum_{k=0}^{s'+1}
        \tilde \alpha_{k} \partial_3^{s'+1-k} \la \underline {\phi \phi^* } j_k   \ra
        }}
        This \Ward
     identity is very similar, in structure,
      to the one we considered in \wards . Namely, the last  term
  contains a sum over various currents which give rise to different
     functional forms. Therefore all $\tilde \alpha_k$'s are fixed. $\tilde \alpha_{s'}$ is
      nonzero  because $[Q_{s'}, j_2]$ contains $j_{s'}$, \qss . The rest of the terms are fixed to the same coefficients that we would have if
     we were considering the same \Ward identity in a free theory of a complex boson\foot{
     In \wardint\ all $\alpha_k$ with even $k$ are set to zero.}.
      In particular, as shown in appendix J, $[Q_{s'},j_2]$ produce a $j_1$ whose overlap with
      $B'$ is nonzero.
  Once we have shown that $j_1$ appears in $B'$, we can
    consider the \Ward identity for $Q_{s}$ on $\langle B j_1 \rangle$.
   \eqn\wardi{
   0 =\langle [Q_s , B j_1 ] \rangle = ( \partial_1^{s-1} - \partial_2^{s-1}) \langle B' j_1 \rangle +
   \langle B [ Q_s , j_1] \rangle
   }
Here the first term is non-zero.
  Analyzing  this \Ward identity,   one can show that
   it can be obeyed, for a non-zero first term, only if $[Q_{s} , j_{1}]$
   contains a $j_s$ in the right hand side. The analysis of this \Ward identity is somewhat
   similar to the previous ones and it is discussed in more detail in appendix J.
    Here we have used that $Q_{s}$ is a current
   that appears in the right hand side of $\underline{j_2 j_2}_b$.
   In other words, $Q_s$ is built out of the same current $j_s$ that appears in the right hand
   side of $[Q_s , j_1]$. This implies that $\langle j_s j_s j_1 \rangle $ would need to be
   non-zero\foot{Let's emphasize again we consider $j_s$ that appears in the OPE of $j_2 j_2 \sim j_s$. Then, in principle, we can have $[Q_s , j_1] = \alpha j_s + \beta \tilde{j_s} +...$. The \Ward identity \wardi\ implies that $\alpha$ is non-zero. This implies that $\la j_s j_s j_1 \ra$ must be non-zero.}. However all such three point functions are zero when the two $j_s$ currents
   are identical \basicfact .
   Since we reached a contradiction, we conclude that $B'$ is zero.

   We would like now to show that $\tilde B$ is the same as $B$.  We know that $\tilde B$ is non-zero because we can
   consider the \Ward identity for $Q_s$  on $\langle B j_2 \rangle$, and
    use that $j_s$ appears in the right hand side of $[Q_s, j_2]$.
    This implies that
  $\langle \tilde B j_2 \rangle$ is nonzero. In fact, we can normalize
   it in such a way that $j_2$ appears in the same way on $B$ and $\tilde B$.
   Then we can consider $B - \tilde B$.
    This does not contain a $j_2$. Here we are using that there is a unique $j_2$ in the theory.
   Say that $j_{s'}$, with $s'$ even and $s'>2$,
    is a candidate current to appear in the right hand side of $B-\tilde B$.
   We can consider
   the $Q_{s'}$ \Ward identity for $ \langle (B- \tilde B) j_2 \rangle $.
   This \Ward identity has a form similar to \wardint ,
   \eqn\wardfi{
 0 =      \gamma ( \partial_1^{s'-1} +   \partial_2^{s'-1}  ) \langle \phi \phi^* j_2 \rangle +
  \sum_{k =0}^{ s'+1}  \tilde \alpha_k   \partial_3^{ s' +1 - k }
 \langle \phi \phi^*  j_k \rangle
 }
  Here we are assuming that $\tilde \alpha_{s'}$ is non-zero. However, this equation would also
  show that $\tilde \alpha_2$ is non-zero which is now in contradiction with the fact that $B - \tilde B$
does not contain $j_2$  (see appendix J).
  So we have shown that $B-\tilde B$ cannot have any current of spin $s'>0$.

Let us now focus on a possible spin zero operator $j'_0$. We put a prime to distinguish it from $j_0$ which is the
  one that appears in $B$.  $j_0'$ might or might not be equal to $j_0$. If $ \la j_0 j_0 ' \ra \not = 0$,
then we can consider the \Ward identity for $Q_4$ acting on $\la (B-\tilde{B}) j_0 \ra$.  Using \qfourz\ we
get a non-zero term when $Q_4$ acts on $j_0$ producing $j_0$. However, this \Ward identity
 cannot be obeyed
if we are setting the term involving $j_2$ to zero. Thus,  $\la j_0 j'_0 \ra =0$.  Then
   there is some even current in the  $\underline{j_2 j_2}_b$, call it $s''$,
   such that $[Q_s, j_{s''}] \sim j'_0 + \cdots $. From \gen\ this implies that $s'' < s$.
   Then we consider the action
   of   $Q_s$ on $\langle (B -
   \tilde B ) j_{s''} \rangle $. This action produces (up to derivatives)
   both $\langle (B -
   \tilde B ) j'_0 \rangle $ and $\langle (B -
   \tilde B ) j_2\rangle $. But given that the second is zero, then the first is also zero.
  This \Ward identity is very similar to the others we have been discussing (see appendix J). Here it is important
  that $s'' < s$, so that we get a constraint on $\langle (B -
   \tilde B ) j'_0 \rangle $. So we conclude that there cannot be any $j_0'$ in $B- \tilde B$.

   In conclusion, we have shown that \propeqn\ holds when $x_1$ and $x_2$ are null separated along
   the minus direction.

  Now, once we argued that  \propeqn\ is true, then we can consider any $n$ point function
  of bilinears $\langle B(\underline{x_1 , x_2}) \cdots B(\underline{x_{ 2n-1},x_{ 2 n}}) \rangle$.
  We have an infinite number of constraints from all the conserved charges \propeqn . These
  constraints take the form
  \eqn\constrin{
  \sum_{i=1}^{2 n} \partial_i^{s-1} \langle B(\underline{x_1 , x_2}) \cdots B(\underline{x_{ 2n-1},x_{ 2 n}}) \rangle =0 ~,~~~~~~~~~s =2,4,6, \cdots
  }
  where $\partial_i = \partial_{x_i^-}$. This constrains the
   $x_i^-$ dependence of this correlator. We show in appendix E that this
   constrains the correlator to be   a sum of functions of
  differences $x_i^- - x^-_j$, $x^-_l- x^-_k$. More explicitly,   the $x_i^-$ dependence
  is such that the correlator is a sum of functions of the form $\sum_\sigma g_\sigma (
  x^-_{\sigma(1)} - x^-_{\sigma(2) } , \cdots , x^-_{\sigma(2n-1)} - x^-_{ \sigma(2 n) } )$, where
  $\sigma$ are the various permutations of $2n$ elements. In principle, these functions, $g_\sigma$ are
  all different.

  In addition, we know that the correlator should respect conformal symmetry and rotational invariance. Thus, it should be a function of conformal dimension $1/2$ at each location and a
  function of distances $d_{ik}$. Of course, $d_{ij}^2 = x_{ij}^+ x_{ij}^- + ( y_i - y_j)^2 $.
  Thus, we can now write the function in terms of distances. Let us explain this point a bit more. We have defined the bilocals in terms of an operator product expansion. But we have also noticed that we can also view these bilocals as special conformal block like objects
  written in terms of currents. From that perspective, the transformation properties under
  conformal transformations are identical to those of a product of bosonic fields. This is what
  we are using at this point.
  Thus, for each function $g_\sigma$, say $g(x_{13}^-, x_{24}^-, \cdots )$ we can now write
  $ \hat g( d_{13} , d_{24} , \cdots )   $.
  But in addition, under conformal transformations, the function has to have weight one half
   with respect to each variable.
  Thus, it can only be a function of the form
  \eqn\functg{
  \hat g( d_{13} , d_{24} , \cdots ) = { 1 \over d_{13} } { 1 \over d_{24} }\times ( { \cdots } )
  }
  where the product runs over $n$ distinct pairs of distances and each point $i$ appears in
  one and only one of the terms. A distance  between two points in the same bilocal is zero and
cannot appear. So such terms
  are not present.
  In addition, the permutation symmetry under exchanges of $B$'s and also of the two arguments
  of $B$ imply that the overall coefficients of all terms
  are  the same, up to disconnected terms, which are given by lower point functions of $B$'s
  by cluster decomposition.
   For example, for a two point function of $B$'s we have
   \eqn\correlat{
   \langle B( \underline { x_1 , x_2} )   B( \underline { x_3,x_4}
   ) \rangle = \tilde N  \left( { 1 \over d_{13} d_{24} } + { 1 \over { d_{14} d_{23} } } \right)
   }
 The overall coefficients of the connected $n$ point function of $B$'s can be determined from the
 one in the $n-1$ point function by expanding one of the $B$'s and looking at the stress
 tensor contribution, which is fixed by the stress tensor \Ward identity.
Therefore, all $n$ point functions of $B$'s are fixed up to a single  constant, the constant in
the two point function of the stress tensor, which is (up to a numerical factor)
 the same as $ \tilde N  $ in \correlat .

We can  consider a theory of $N$ free bosons, with
 \eqn\defb{
  B(\underline { x_1 , x_2 }  ) =
   \sum_{i=1}^N  \underline {:   \phi_{i} (x_1) \phi_{i} (x_2) : }
  }
 where the $::$ imply that we do not allow contractions between these two $\phi$ fields.
This theory has a higher spin symmetry and
we also get the same correlators, except that the constant in front of the stress tensor and \correlat \ is
$\tilde N = N$.
Another way to state the result is that all the correlators of the currents are the same as the
ones we have in a theory of $N$ bosons with $N$ analytically continued, $N \to \tilde N$.
However, as we argue below $\tilde N $ should be an integer.

We did not show
  that the operators for the elementary fields (or partons) are present as good operators
  in the theory. It is clear that one cannot show that by looking purely at correlators
  on the plane, since one can project them out by imposing an $SO(N)$ singlet condition which
  would leave all remaining correlators obeying the crossing symmetry relations, etc.

  Let us discuss further  lessons. First, we observe that the expansion of $B$'s
  contains the currents with minus indices $j_s$. Thus, the three point function of $B$'s contain
  the three point functions of currents with minus indices. All of these are the same
   as those of the free boson theory. Note that the possibility of odd three point functions has
   disappeared. Thus, we never needed to know much about the nature of the odd structures for the
   three point functions.
  From the OPE of two currents with minus indices
   we can get currents with arbitrary indices. Thus, from the six point function of $B$'s we
   get the three point function of the currents for arbitrary indices, which coincides with
   the free boson answer. Similarly from a $2 n$ point function of $B$'s we get an $n$ point
   function of currents with arbitrary indices which is equal to the free boson one for a
   theory of $N$ bosons. By further performing OPE's we get other operators such as
 ``double trace'' or ``double sum'' operators.   In conclusion, we have fixed all the
 correlation functions of the stress tensor, and also all the correlation function for all
 the higher spin conserved currents that appear in the $j_2 j_2$ operator product expansion.
 These are currents that have even spins. In addition, we have also fixed the correlators of
 all other operators that appear in the operator product expansion of these operators.
 All such correlators are given by the corresponding correlators in a theory of $N$ free
 bosons restricted to the $O(N)$ invariant subsector.

We should remark that, though
 we made statements regarding currents that can be written as $O(N)$ invariant
bilinears,
this does not mean that the theory is the ``free $O(N)$'' model.
The theory can contain
additional conserved currents (and still only one stress tensor).
 For example, for $N= 2 M$, we can consider
$M$ complex fields and restrict to the $U(M)$ invariant sector. This theory obeys all the assumptions
of our theorem. In particular, it still has a single conserved spin two current.
 However,  we also  have additional currents
of odd spin. Of course, the currents with even spin that appear in the $j_2 j_2$ OPE
are still given by an $O(2 M) $ invariant combination
of the fields and have the same correlators that we discussed above.
 Probably a little more work
would show that if we had odd spin currents, they should also behave like those of free fields. Presumably
one would construct a bilocal operator from the odd currents and argue as we did above.
One could also wonder about other even spin currents which do not appear in the $j_2 j_2$ OPE.
Probably there cannot be such currents (with a single conserved spin two current),
but we did not prove it.

 Note that the correlation functions of stress tensors are all equal to the ones in the
free theory. In particular, if we created a state with an insertion of the stress tensor at the origin
we could compute the energy correlation functions that would be measured by idealized detectors
(or calorimeters) at infinity. These are the energy correlation functions considered, for example in \refs{\BashamIQ,\HofmanAR}, and are computed by particular limits and integrals of correlation functions of the
stress tensor. The $n$ point energy correlator for a state created by the
stress tensor is schematically $\la 0| T^\dagger(q) \epsilon(\theta_1) \cdots \epsilon(\theta_n) T(q) |0\ra $, where $T(q)$ is the insertion of a stress tensor
 operator of four momentum (roughly) $q$ at
the origin and $\epsilon(\theta)$ are the energies per unit angle
 collected at    ideal calorimeters sitting
 at infinity   at the angle $\theta$. These
will give the same result as in the free theory. Namely, that the energy is deposited in two localized points,
corresponding to the two partons hitting the calorimeters.  This result is qualitatively similar
to the Coleman-Mandula result for the triviality of the S-matrix.
These energy correlation functions are infrared safe (or well-defined) observables which are,
conceptually,  rather
close  to the S-matrix. Here we see that these energy distributions are trivial.

 \subsec{Quantization of $\tilde N$: the case of bosons }

The basic idea for showing that $\tilde N$ is quantized uses the fact that the $N =\infty$
theory and the finite $N$ theory have a different operator spectrum. The finite $N$ spectrum
is a truncation of the infinite $N$ spectrum. This point was also emphasized recently in
 \ShenkerZF .

 In order to prove the quantization of $\tilde N$ we argue as follows.
In the theory of $N$ bosons we consider the operator
\eqn\operabo{
{\cal O}_q =  \delta^{[i_1, \cdots , i_q ]}_{[j_1, \cdots, j_q ] } ( \phi^{i_1} \partial \phi^{i_2} \partial^2 \phi^{i_3}
\cdots \partial^{q-1} \phi^{i_q} ) ( \phi^{j_1} \partial \phi^{j_2} \partial^2 \phi^{j_3}
\cdots \partial^{q-1} \phi^{j_q} )
}
where the $\delta$ function is the totally antisymmetric delta function of $q$ indices. It is
the object that arises we consider a contraction of two $\epsilon $ tensors of the from
\eqn\constr{  \delta^{[i_1, \cdots , i_q ]}_{[j_1, \cdots, j_q ] }  \propto \epsilon^{i_1, \dots , i_q , i_{q+1} , \cdots , i_N} \epsilon_{j_1, \cdots ,j_q, i_{q+1} \cdots i_N}
}
This operator, \operabo , can be rewritten as sum of products of $q$ bilinear operators.
Once we have written it as a particular combination of bilinear operators we can consider any value of
$N$ and we can imagine doing analytic continuation in $N\to \tilde N$, with $q$ fixed.

In particular, we can consider the norm of this state. We are interested in the $\tilde N$
dependence of the norm of this state.
We can show that
\eqn\normsta{
 \la {\cal O}_q {\cal O}_q \ra = \tilde N ( \tilde N -1) ( \tilde N -2 ) \cdots ( \tilde N - (q-1) )
}
where we only indicated the $\tilde N$ dependence. Since we have $q$ bilinears, the
series expansion in $1/\tilde N$ has only $q$ terms.  In addition, the result should vanish
for $\tilde N =1,2, \cdots , q-1$, and the leading power should be $\tilde N^q$.

Now, imagine that $\tilde N$ was not integer. Then we could consider this operator for
$q= [ \tilde N] +2$, where $[\tilde N]$ is the integer part of $\tilde N$.
Then we find that \normsta\ is
\eqn\nofs{
\la{\cal O}_ {[ \tilde N] +2} {\cal O}_{[ \tilde N] +2}
 \ra = ( {\rm positive} ) ( \tilde N -  [\tilde N ]  - 1)
}
where we only wrote the last term in \normsta , which is the only negative one.
Thus, we have a negative  norm state unless $\tilde N$ is an integer. Therefore, unitarity
forces $\tilde N$ to be integer. Here we have phrased the argument in terms of the norm of
a particular state, which might be changed by choosing a different normalization constant.
 However, we can get the same argument by consider the contribution
to the OPE of a state like ${\cal O}_q$. If the norm is negative, then we will get a negative
contribution to the OPE in the channel that is selecting this particular ${\cal O}_q$.

   \subsec{Fermionic-like bilocal operators}

We now return to the second case in \casest , where we have to use \correcf .
  We can now repeat the operations we did for the bosonic case in order to argue that
  \eqn\conffu{
  [ Q_s , F_{-}(\underline{x_1,x_2}) ] = ( \partial_1^{s-1} + \partial_2^{s-1} ) F_{-}(\underline{x_1,x_2})
  }
The arguments are completely similar to the case of the boson. One needs
to apply the same \Ward identities. All the arguments are very similar,
except that we now take the fermion like limit  \limitsdef\ and use
the functions in \limitfcorr . One can run over all the arguments presented for the boson-like
limit
 and one can check that all the \Ward identities  have the same implications
for the fermion-like case. This is shown in appendix J.

 Once we show \conffu , we can now constrain the form of any correlator of the form
 $ \langle F_-(\underline{x_1,y_1} ) \cdots F_-(\underline{x_n,y_n}) \rangle$.
  Again it involves functions of differences of $x_{ij}^-$ with each $i$ appearing only in
  one argument of the function.
 In addition, we know that they should be rotational invariant and conformal covariant.
 However, as opposed to the bosonic case,
  in this case the conformal transformations of $F_-$ are those of a product of
 fermions.
 We can take into account these transformations by using factors of fermion propagators,
 $x_{ij}^+/d_{ij}^3 $. Together with permutation (anti) symmetry
 these  constraints imply that the correlators are those of free fermions.
  For example, for the four point function we get
 \eqn\concl{
 \langle F_-(\underline{x_1, x_2}) F_-(\underline{x_3 , x_4} ) \rangle = \tilde N \left( { x^+_{13} \over d_{13}^3 }
 { x^+_{24} \over d_{24}^3 } - { x^+_{14} \over d_{14}^3 }
 { x^+_{23}  \over d_{23}^3 }  \right)
 }
Similarly we can use the symmetries to fix all  $n$ point functions of $F$ in terms of the single
parameter  $\tilde N$ that appears in \concl , or in the two point function of the stress tensor.
For $\tilde N = N$, these $n$ point functions agree with the ones we would obtain in a theory of $N$
Majorana fermions with
\eqn\defin{
F_-(\underline{x_1, x_2}) = \sum_{i=1}^N \underline{: \psi^i_{-}(x_1) \psi_{-}^i(x_2) :}
}

The expansion of $F_-$ contains all currents of twist one with minus indices.
The further OPE of such currents contains twist one currents with other indices and also
a scalar operator of twist two.
In a theory of free fermions this is
\eqn\jzerof{
\tilde j_0 = \sum_{i=1}^N \psi_{+}^i \psi_{-}^i
}
Thus, all the correlators of currents of even spin, plus the twist two scalar operator are fixed by
the higher spin symmetry to be the same as in the theory of $N$ free fermions.

\subsec{Quantization of $\tilde N$: the case of fermions}

One can wonder whether there exists any theory $\tilde N$ is non-integer. Since $\tilde N $ appears
in the two point function of the stress tensor, we know that $\tilde N > 0$.
We now argue for the quantization of $\tilde N$. The argument follows from considering
the operator
\eqn\operc{
{ \cal O}_q  \equiv  :  ( \tilde j_0 )^q :
}
For general $\tilde N$, this operator is  defined by looking at the appropriate term in the operator product expansion of $q$ $\tilde j_0$
operators that are coming together. And $\tilde j_0$ is itself also defined via a suitable operator product
expansion of the original bilocals $F_-$.
We can now compute the two point function of such operators, \operc .
More precisely, since the two point
function is arbitrarily defined, we can look at the contribution to the OPE of the exchange of the
operator ${\cal O}_q$.  By the way, notice that the correlation function of any of the current bilinears
or any product of such bilinears, is given by making an analytic continuation in $N \to \tilde N$ of
the free fermion results. It is important that when we make this analytic continuation, the number of
currents in a correlator should be independent of $N$.

A crucial observation is that
 in a free fermion theory with $N$ fermions we have that \operc\ is zero for $q = N+1$. This
means that such a contribution should not be present in the OPE.

Now, let us fix $q$ and compute  the two point function
of ${\cal O}_q$ for general $\tilde N$. (More precisely, we are talking about the contribution of
this operator to the OPE).  The expression has a $1/\tilde N$ expansion with
precisely $q$ terms, since that is the number of terms we get in a free fermion theory.
 We argue below that it should have the following $\tilde N$ dependence
\eqn\normfo{
\langle {\cal O}_{q} { \cal O}_{\tilde q }  \ra =   \tilde N ( \tilde N -1) ( \tilde N -2) \cdots (\tilde N -(q-1) )
}
First, we see that \normfo\ has precisely $q$ terms. Second, notice that it should vanish for
$\tilde N = 1, 2, \cdots , q-1$, since for such integer values, we have correlators identical to those
of the free fermion theory where ${\cal O}_q$ vanishes.

If $\tilde N$ was not an integer, then we could set $q = [\tilde N] + 2$, where $[\tilde N]$ is the integer
part of $\tilde N$.  Then the norm \normfo\ would
be equal to
\eqn\normop{
\langle { \cal O}_{[\tilde N]+2 }  { \cal O}_{[\tilde N] +2 } \ra  = ( {\rm positive } ) ( \tilde N - [ \tilde N] -1 )
}
where we only wrote the last factor in \normfo . The rest of the factors are positive. However, this
last factor is negative if $\tilde N$ is not an integer. Thus, we see that unitarity forces $\tilde N$ to be
an integer and we get precisely the same values as in the free fermion theory.


\newsec{ Arguments based on more generic three and four point functions }

In this section we derive some of the above results in a conceptually more
straightforward fashion, which ends up being more computationally intensive.
We checked the statements below by using Mathematica.
Of course, if we were to take light-like limits some
of the computations simplify, and we go back to the discussion in the previous section.

\subsec{Basic operations in the space of \Ward identities}

The presence of higher spin symmetries leads to \Ward identities for three point functions
which relate three point functions of conserved currents of different spins. Namely, we start from a three
 point function and demand that it is annihilated by $Q_s$.
 This imposes interesting
constraints for the following reason.
The three point functions have a very special form due
 to conformal symmetry and current conservation. The action of the higher spin charge
  \gen\  gives a linear combination of these three point functions and their derivatives.
  These derivatives do not commute with the action of the conformal group, so this single
  equation is equivalent to larger set of equations constraining the coefficients of the various three point functions. In other words, since conformal symmetry restricts the functional
  form of the three point functions, the single equation that results from $Q_s$ charge conservation is typically enough to fix all the relative coefficients of the various
  three point functions that appear after we act with $Q_s$ on each of the currents.

Let us first describe some operations that we can use over and over again to constrain
the action of the symmetries.

Imagine that we are trying to constrain the action of $Q_s$ on a current $j_x$ and we
would like to know whether current $j_y$ is  present or not  in the transformation law
\eqn\transform{
[Q_4 , j_x] =  j_y + ...
}
(generically, with some derivatives acting on $j_y$). Recall that through \jxjy , then
$[Q_4 , j_y ] = j_x + \cdots$.
Now the basic \Ward identity we can use is the one resulting from the action of
 $Q_4$ on   $\la j_2 j_x  j_y \ra$. First of all, notice that from the variation of $j_2$ we will necessarily get (see \qss )
 the term
$\la j_4 j_x j_y \ra$ which must be non-zero if the $j_x$ and $j_y$ appear in each other transformations under $Q_4$.
The simplest possibility would be to find that the only solutions of the \Ward identity
 are such that $\la j_4 j_x j_y \ra = 0$. Thus, our assumption about the presence of $j_y$ in the variation of $j_x$ was wrong.
This is {\rm the basic operation of the elimination} of $j_y$ from $[Q_4,j_x]$.

If we can find that  solutions of the \Ward identity with $\la j_4 j_x j_y \ra \neq 0$ exist, this is consistent with the presence of $j_y$ in the variation of $j_x$ but does not necessarily
imply it.

Another basic \Ward identity operation can be used to
the check that $j_y$ {\rm is definitely  present} in the transformation of $j_x$. This is
done via the \Ward identity resulting from the action of $Q_4$ on
 $\la j_4 j_x  j_x \ra$ \foot{In practice,   we first check
 that $[Q_4, j_2] \sim \pa^3 j_2 + \cdots $ and then we use $\la j_2 j_x  j_x \ra$.}.
  Notice that from the transformation of $j_4$ we will necessarily (see \qss , \jxjy )
  get the term
$\la j_2 j_x j_x \ra$ which must be non-zero due to the fact that $j_2$ generates conformal
transformations.
 Now imagine that for the
solution of the \Ward identity  to exist the term $\la j_4 j_x j_y \ra$
should be necessarily non-zero. This means that $j_y$ is necessarily
present in the transformation of $j_x$.

Very often using these two operations allow us  to fix completely which  operators
do appear in the transformation of the given conserved current.

\subsec{From  spin three  to spin four}

In this section we  show that in any theory that contains a conserved current of
spin three then  the conserved spin four current is necessarily present.

Let us first consider the most general CFT that has spin three current.
We have that the most general transformation has the form
\eqn\mostgen{\eqalign{
[Q_{3}, j_{2}] &= \alpha_{0} \pa^4 j_{0} + \alpha_{1} \pa^3 j_{1} + \alpha_{2} \pa^2  {j}_{2} +\alpha_{3} \pa j_{3}+\alpha_{4}  {j'}_{4},\cr
[Q_{3}, j_{3}] &= \beta_{1} \pa^4  { j'}_{1} + \beta_{2} \pa^3 j_{2} +\beta_{3} \pa^2 {j'}_3 + \beta_{4} \pa j_4 + \beta_{5} j_5.
}}

Let's make several comments on this expression.
Primes stand for the fact that   the same or a different  current of the same spin
 can, in principle, appear in the right hand side.
 Notice that $\alpha_{4} =0$ since
otherwise $Q_{3}$  is not translation invariant. Integrating both sides we would get
$[Q_3, P_-] = Q_4 $, which would mean that $Q_3$ is not translation invariant, in contradiction with
\chargas\ and current conservation.
From \qss ,   $\alpha_{3}$ should be non-zero, and so is $\beta_{2}$.

As the next step we consider the \Ward identity obtained by $Q_3$ acting on $\la j_2 j_2 j_3 \ra $
\eqn\WIgen{
0 =   \la[Q_{3},j_2 (x_1) j_2 (x_2) j_3 (x_3)] \ra = 0
}
This is essentially the same \Ward identity we considered in section five.
The solution exists only for $\beta_4 \neq 0$. In other words, the spin four current
is necessary present in the theory.

Another feature of this exercise that is worth mentioning is that the general solution of the \Ward identities  involves three distinct pieces
\eqn\gensol{
\la j j j \ra = \la j j j \ra_{boson} + \la j j j \ra_{fermion} + \la j j j \ra_{odd}
}
the first two pieces correspond to the free boson and free fermion three point functions respectively. We expect these solutions to be present
in the theory of complex boson and fermion.
The odd piece does not come from any of free theories and is parity
violating \foot{Since it gives odd contribution to the three point function of stress tensors.}. We will elaborate the nature of the odd piece later.
But we should emphasize that at the level of three point functions one can find three independent solutions of higher spin \Ward identities \foot{We will show below that the odd solutions appear in the theories with higher spin symmetry broken at ${1 \over N}$ order .}.

After an  illustration of this particular example we can formulate the general recipe.
Assume that the theory has a conserved current $j_{s}$  of spin $s \neq 4$ and $s>2$. We again
consider the \Ward identity obtained by $Q_{s}$ acting on $\la j_2 j_2 j_{s} \ra$ and arrive at the conclusion
that $j_4$ must be present in the spectrum. This type of  \Ward identity is especially simple to analyze using the light cone limit
described in the previous section.

So that we take  it as a given that we always have a spin four current.

\subsec{Analysis of three point functions using the spin four current.}

In this section we look in detail at the action of the $Q_4$ charge. Again we present only
the results of the computations which are straightforward but tedious.
We can argue that
\eqn\onthestress{
[Q_{4}, j_2] = \partial^5 j_0 + \partial^3 j_2 + \partial j_4
}
We have eliminated $j_1$ and $j_3$ by  considering  the \Ward identity corresponding to the action of
 $Q_{4}$   on  $\la j_2 j_2 j_1 \ra$ and $\la j_2  j_2 j_3\ra$.
 We used general transformation laws for $j_1$, $j_3$ \gen .
Then we consider the action of
 $Q_{4}$ at $\la j_2  j_2  j_2 \ra$ and use that we have already shown that
  $\la j_2 j_2 j_4 \ra$ is nonzero.
 The stress tensor three point function can have three different pieces
\eqn\pieces{
\la j_2 j_2 j_2\ra = \la j_2 j_2 j_2\ra_{boson} + \la j_2 j_2 j_2\ra_{fermion} +
\la j_2 j_2 j_2 \ra_{odd}
}
The \Ward identity gives three different solutions involving other spins corresponding
to these three different pieces. In other words, the \Ward identity equations for the
boson, fermion and odd pieces do not mix.
However, if the boson or odd pieces in \pieces\ are non-zero then the \Ward identity implies
that the current $j_0$ exists and appears as in \onthestress . The fermion solution does
not require a $j_0$.

  If the stress tensor is unique, then the fact that $j_0$ exists, implies that
  $\langle j_2 j_2 j_2 \ra_{f} =0$. This is done by considering the $Q_4 $ \Ward Identity for
  $\la j_0 j_2 j_2 \ra $. $Q_4$ on $j_0$ gives $j_2$ due to \jxjy\ and \onthestress.
 The non-zero fermion part of these three point functions is not compatible with this \Ward identity.

Then the problem separates into two problems. First we consider the case where
  $\la j_2 j_2 j_2 \ra_f =0$ and then the case when $\la j_2 j_2 j_2 \ra_b =0$.
  Both cannot be zero due the fact that $j_2$ generates conformal transformations for $j_2$, and
  the fact that the odd piece does not contribute to the action of conserved charges\foot{Consider $\int d x^{-} d y \la j_{s_1}(x) j_{s_2}(x_2)  j_{s_3}(x_3) \ra_{odd}$  with $y_2 = y_3 = 0$ then under $y \to - y$ the integrand is odd and, thus, the integral vanishes.}.
 The odd piece can be eliminated at this point by an energy correlation argument, as
 shown in appendix C. But the reader not familiar with that technique can wait a little
 longer until we eliminate it in a more straightforward way.

\subsec{Constraining the four point function of $j_0$}

 First we can constrain the action of $Q_4$ on $j_0$.
 By a method very similar to the one we used for the stress tensor we can prove that
\eqn\scalartrans{
[Q_{4}, j_0 ] =  \partial^3 j_0 +\gamma \partial j_2
}
where we wrote the constant $\gamma$ appearing in the transformation explicitly.
Again we eliminate $j_1$ and $j_3$ by considering the $Q_4$ \Ward identities on
 $\la j_2 j_1 j_0\ra$ and $\la j_2 j_3 j_0\ra$. The other two terms can be found
 from the \Ward identity for $Q_4$ acting on $\la j_0 j_2 j_2 \ra$, and using
 \onthestress\ and the fact that $\la j_0 j_0 j_2 \ra$ is nonzero.

Now we consider $Q_4$ acting on the four point function   $\la j_0 j_0 j_0 j_0 \ra$.
This gives
\eqn\WIsss{
\pa^3_1 \la j_0 j_0 j_0 j_0 \ra + \gamma \pa_1 \la j_2 j_0 j_0 j_0   \ra  + [1\leftrightarrow 2] + [1 \leftrightarrow 3] + [ 1 \leftrightarrow 4] =0
}
In order to solve this we first need to write the most general four point function with one
insertion of the stress tensor and three scalars, $\la j_2 j_0 j_0 j_0 \ra $.
 This can be done using the techniques described in
\refs{\GiombiRZ , \CostaMG} . There are two possible forms, one is parity even and the other is
parity odd. The equation \WIsss\ splits into two one for the parity even and the other
for the parity odd piece.
The general form for the parity even piece
involves certain conformal invariants $Q_{ijk}$
constructed out of the positions and the polarization tensors
\eqn\stresste{\eqalign{
\la j_{2}(x_1) j_0(x_2) j_0(x_3) j_0(x_4)\ra &=  {Q_{123}^2 g (u,v) + Q_{124}^2 g (v,u) + Q_{134}^2 \tilde{g}(u,v) \over x_{12}^2 x_{34}^2} \cr
\tilde{g}(u,v) &= {1 \over u} g ({v \over u},{1 \over u})
}}
where the relevant conformal invariants simplify to
\eqn\confinv{\eqalign{
Q_{i j k} &= {x_{i,j}^{+} \over x_{i,j}^2} - {x_{i,k}^{+} \over x_{i,k}^2} ~,~~~~ u = { x^2_{13} x_{24}^2 \over x^2_{12} x^2_{34} } ~,~~~~~~~~ v =  { x^2_{14} x_{23}^2 \over x^2_{12} x^2_{34} }
}}
because we consider only the minus polarization.
We define  $f(u,v)$ via
\eqn\fourscalar{\eqalign{
\la j_0 j_0 j_0 j_0  \ra &=  {f (u,v) \over x_{12}^2 x_{34}^2} \cr
f  (u,v) &= f  (v,u) = {1 \over v} f ({u \over v}, {1 \over v})
}}
Inserting \fourscalar\ and \stresste\ into \WIsss , we get an equation which depends
both on the cross ratios and the explicit points $x_i$. By applying conformal transformations,
and a lot of algebra,
we can get a set of equations purely in terms $f$, $g$ (and $u$ and $v$).
There is a solution where $g=0$, which has a factorized dependence of the coordinates.
There is also a solution with both $g$ and $f$.
The sum of these two solutions is
\eqn\fgsol{\eqalign{
f  (u,v) = & \alpha (1 + {1 \over u} + {1 \over v}) + \beta ({1 \over \sqrt{u}} + {1 \over \sqrt{v}} + {1 \over \sqrt{u v}}),
\cr
\gamma g(u,v)  = &  \beta  {9 \over 20 \sqrt{u}}.
}}
By taking OPE of $\la j_0 j_0 j_0 j_0\ra$ we can extract, for example, the $\la j_0 j_0 j_0\ra$ structure constant.
We can fix $\gamma$ then by considering the $j_2$ \Ward identity. In other words,
integrating $j_2$ to get the action of  $P_-$   on $\la j_0 j_0 j_0\ra$, for example.

These are the free field theory correlators. The term proportional to $\alpha$ is the disconnected
contraction and the one involving $\beta$ is the connected one. In a theory of $N$ free
scalars we can set $\alpha =1$ by a choice of normalization for the operators.
Then $\beta \sim 1/N$. Notice that we were able to fix two four point functions using just one \Ward identity.

\subsec{No parity odd piece}

While it is clear that we cannot write an  odd piece for the four point function of scalars we
can do it for $ \la j_2 j_0 j_0 j_0   \ra$.
 The unique structure in this case takes the following
form in the embedding formalism  (see \CostaMG\ for conventions)
\eqn\oddbosonstr{\eqalign{
\la j_2 (x_1)   j_0 (x_2)  j_0 (x_3)  j_0 (x_4) \ra &\sim {\eps (Z_1, P_1, P_2, P_3, P_4) \over (P_1 P_2)^2 (P_1 P_3)^{3/2} (P_1 P_4)^{3/2} (P_3 P_4)^{1/2}} \cr
&\left[Q_{123} g_1 (u,v) + Q_{134} g_2 (u,v) + Q_{142} g_3 (u,v)  \right] \cr
}}
Inserting this in the \Ward identity \WIsss,
 $\sum \pa \la j_2 j_0 j_0 j_0 \ra = 0$ we find that there is no solution.

By taking the OPE of $\la j_2 j_0 j_0 j_0  \ra$  we can concentrate on the twist one channel where the stress tensor is propagating.
The relevant three point functions are $\la j_2 j_0 j_2  \ra$ and $\la j_2 j_0 j_0    \ra$.

Both of them are non-zero. The function $\la j_2 j_0 j_0\ra$ is non-zero because stress tensor generates conformal transformation
and  $\la j_2 j_0 j_2\ra$ is non-zero because of the $Q_4$ \Ward identity  for $\la j_2 j_2 j_2 \ra$.

Due to the triangle inequality \trianglerule\  the $\la j_2 j_0 j_2 \ra$ is the only odd three point function that is non-zero in the twist one sector OPE expansion of $\la j_2 j_0 j_0 j_0  \ra$.

The fact that the four point function does not have the odd piece forces
 us to set the odd piece of $\la j_2 j_0 j_2 \ra$ to zero. Then through the higher spin \Ward identities we will set the odd part of the whole tower of conserved currents three point functions to zero. Thus, while the odd pieces
of three point functions respect higher spin symmetry, at the level of four point functions they are eliminated in the theories where
higher spin symmetry is exact.

\subsec{Case of fermions}

The above discussion can be repeated for the case that $\la j_2 j_2 j_2 \ra_f \not =0$.
In this case we do not expect a twist one, spin zero field.
Here we can show that
\eqn\sofe{
[ Q_4 , j_2] = \partial^3 j_2 + \partial j_ 4
}
In principle, we could act with $Q_4$ on the four point function
$ \la j_2 j_2 j_2 j_2 \ra $ and use \sofe\ to fix it completely. Though this probably
works, we have not managed to do it due to the large number of conformal structures that
are possible.

Instead, one can take a longer route by first showing that a certain twist two scalar
operator $\tilde j_0$ exists, finding its transformation laws under $Q_4$ and then
showing that its four point function is the same as that of $\tilde j_0 = \epsilon^{\alpha
\beta } \psi_{\alpha } \psi_{\beta}$ in a theory of free fermions.

In this section we consider not just the minus component of the currents but also
the third, or perpendicular component.
For the charge $Q_4$ we continue to focus on the all minus component.

We consider the \Ward identity from the action of $Q_4 $ on
 $\la j_{2} j_{2} j_{2-\perp} \ra$.

The new ingredient is the variation $[Q_{4}, j_{- \perp}]$ which should have  twist two operators in the right hand side.
We write the most general form of this variation  denoting by $\tilde{j}_s$  a twist two
operator of spin $s$ (and
without tilde for the twist one ones)

\eqn\varrrB{\eqalign{
[Q_{4}, j_{2\perp}] &= \sum_{i=0}^4 \pa^{4-i} (\tilde{j}_i +\pa_{\perp} j_i + j_{(i+1) \perp})
}}
where all implicit tensor indices are minus.

Now as we are having in mind the $Q_4$
\Ward identity for $\la j_{2} j_{2} j_{2\perp}\ra$ we can ignore all the odd spin twist one current due to \basicfact.  Moreover, we can fix $\tilde{j}_{4-}$ to zero using the translation invariance argument.
 Also we know that for the fermion solution $j_0$ term is absent. And here we are interested in the fermion part of
the solution for $\la j_2 j_2 j_2\ra$.
One can be confused by the lack of perpendicular index in the twist two sector. It comes from
an $\epsilon$ tensor as  $
\eps_{- \perp \mu} \pa^{\mu} = \eps_{- \perp +}   \pa_{-}
$,
so that with more general indices we have
\eqn\symm{
[Q_{4}, j_{2 \mu \nu}] = \pa^2_{-} [\pa_{\mu} \eps_{- \nu \sigma} + \pa_{\nu} \eps_{- \mu \sigma}] \pa^{\sigma} \tilde{j}_0 + ...
}
This  epsilon tensor implies  that in the \Ward identity
  $\la j_{2} j_{2} j_{2-\perp}\ra_{even}$
  should cancel the odd contribution  $\la j_{2} j_{2} \tilde{j}_0 \ra_{odd}$.
 The relevant terms that we need for the \Ward identity are
\eqn\varB{\eqalign{
[Q_{4}, j_{2 \perp}] &= \pa^4 \tilde{j}_0  + \pa^2  \tilde {j}_{2} + \pa^2 \pa_{\perp}  {j}_{2} + \pa_{\perp} j_{4}  \cr
&+ \pa^3  {j}_{2\perp}   + \pa j_{4 \perp} + ...
}}
The dots denote terms that do not have any overlap with $j_2 j_2$.
So we end up with the following \Ward identity to be checked. Here we think about all even parts as being generated by fermion three point functions

\eqn\WItocheck{\eqalign{
&\la [ \pa^3 j_{2}]  j_{2} j_{2\perp}\ra_{even} + \la [ \pa j_{4} ]  j_{2}  j_{2\perp}\ra_{even} + [1 \leftrightarrow 2] = \la j_{2}  j_{2} [ \pa^4 \tilde{j}_{0} ] \ra_{odd}\cr
& + \la j_{2}  j_{2} [ \pa^2_{} \tilde{j}_{2} ] \ra_{odd}+\la j_{2}  j_{2} [ \pa^2 \pa_{(-} j_{2\perp)}] \ra_{even} +\la j_{2} j_{2}  [ \pa_{(-} j_{4\perp)} ] \ra_{even} .
}}

The only new three point functions involved are $\la j_{2}j_{2}\tilde{j}_0\ra_{odd}$ and $\la j_{2}j_{2}\tilde{j}_2 \ra_{odd}$.
Both are fixed by conformal symmetry, and current conservation for $j_2$, up to a constant.
The \Ward identity implies that the twist two scalar
$\tilde j_0$ needs to exist and it needs to appear
in the right hand side of
\eqn\varC{\eqalign{
[Q_{4}, j_{2 \perp}] &= \pa^4 \tilde{j}_0 + \pa^3 j_{2\perp}   + \pa j_{4 \perp} + ...
}}

By performing an analysis similar to what was done for the bosonic case one can show
that nothing else can appear in the right hand side of \varC .
In addition, one can then constrain the action of $Q_4$ on $\tilde j_0$ and find
\eqn\finda{
[ Q_4 , \tilde j_0 ] = \partial^3 \tilde j_0  +  \partial \partial_{[-}  j_{2 \perp]}
}
 This again requires looking at many \Ward identities to eliminate all the other possible
 terms that could appear. Again, notice that the second term involves an $\epsilon$ tensor.

 After this, one can look at the \Ward identity for the four point function
 $\langle \tilde j_0 \tilde j_0 \tilde j_0 \tilde j_0 \ra$. We will need
 $\langle  J_{2 \mu \nu} \tilde j_0 \tilde j_0 \tilde j_0 \ra$, actually,
 the parity odd version of this correlator,  which has a structure similar to \oddbosonstr\ except
 for the different conformal dimension of $\tilde j_0$ (now two).
 After a lot of algebra one can show that this four point function has the form
 of free fermions (see Appendix G).

\newsec{Higher spin symmetries broken at order  ${1 \over N}$  }

So far we considered the case where the symmetries are
exactly conserved. There are some interesting
 theories, mainly large $N$ $O(N), Sp(N),  U(N) $ vector
models where the symmetries are ``almost'' conserved. By this we mean that the anomalous
dimensions of the currents are of order $1/N$. Furthermore the correlators of the theory
have $1/N$ expansion. We will make now some remarks about this case here.

To explain this imagine a situation when the higher spin currents have anomalous dimension $\Delta = s + 1 + O({1 \over N})$. The operators have a single ``trace''  vs. a multitrace structure, and to leading
order in $N$ the correlators of multitrace operators factorize\foot{We should rather speak of a ``single sum''
vs. a ``multi-sum'' structure.}.

Rather than giving a general discussion, here we will focus on one specific case that will illustrate the
method, leaving a more general discussion for the future.

We can illustrate the method by considering a large $N$  $SO(N)$ Chern-Simons matter theory with
an $SO(N)$ level $k$
Chern-Simons action coupled to $N$ Majorana fermions. This theory was considered recently
in \GiombiKC \foot{Actually, they considered a $U(N)$ gauge group, but the story is very similar.
For a similar theory with scalars see \AharonyJZ. } . In this theory, in addition, we have  a coupling $\lambda \sim N/k$ which is very small
when the Chern-Simons level is very large, $k \gg N$. So we now have this second expansion parameter
which we will also use.

In the limit that $\lambda =0$ we have the theory of free fermions. The ``single trace'' operators
in this theory are spanned by the twist one  currents $j_s$ and the twist two pseudoscalar operator
$\tilde j_0$. Let us normalize them so that their two point functions are one.
 Let us consider the most general form of the divergence of the $J_4$ current in this theory
\eqn\wiscalar{
\pa_{\mu} J^{\mu}_{\ \ ---} =  {1 \over \sqrt{N}}  \left[
 a_1 \pa_{-} \tilde{j}_{0} j_{2}+ a_2   \tilde{j}_{0}  \pa_{-} j_{2}  \right].
}
The divergence should be twist three operator with spin three. The terms in the right hand side are
all the operators we can write down in this theory. Note that there are no single trace primaries that
can appear.  Notice that at this point we do not use any information
about the microscopic theory except the spectrum at $N = \infty$.

We can now consider partially broken \Ward identities, by integrating an insertion of \wiscalar\ in a
correlation function
\eqn\corrfu{
  \int d^3 x \la \partial_\mu J^\mu_{~ - - - } {\cal O}_1  \cdots {\cal O}_n \ra =
\sum_{i} \int_{S_i } \la j^n_{\ \ ---} {\cal O}_1  \cdots {\cal O}_n \ra
}
Here we have used Stoke's theorem to integrate the divergence on a region which is the full $R^3$
  minus
little spheres around the insertion of each operator. The surface terms have the usual expression for
the charges in terms of the currents. Except that now they are not conserved. However, since the
non-conservation is a small correction, we can act on each of the operators by the charges, up
to $1/N$ corrections. In the left hand side we can insert the expression \wiscalar\ and use the leading
$N$, factorized correlator. In our normalization the action of the ``charges'' , have a $1/\sqrt{N}$, so
that left hand side and right hand side of \corrfu are of the same order.

As a first example, consider the case where the operators are simply ${\cal O}_1 {\cal O}_2 =
\tilde j_0  j_2 $. In this case, the charges, to leading order annihilate the correlator in a trivial
way, using
\finda, \sofe . All that remains is the integral of the right hand side of \wiscalar . In order for this to vanish,
we need that
\eqn\condi{
a_2  = - { 2 \over 5  } a_1.
}
So this relative coefficient is fixed in this simple way, for all $\lambda$, to leading order in $1/N$.
This is a somewhat trivial result since it also follows from demanding that the special conformal generator
$K^-$ annihilates the right hand side of \wiscalar . We have spelled it out in order
 to illustrate the use of the broken symmetry.

As a less trivial example, consider   the insertion of the same broken \Ward identity in
the three point function of the stress tensor. We will do this to leading order in $\lambda$.
 We get
\eqn\WIcons{
\sum_i  \int_{S_i}  \la j^{n}_{---} j_2 (x_1) j_2 (x_2) j_2 (x_3)\ra \sim  {a_1 \over \sqrt{N}}\int d^3x
\la [  \pa \tilde{j}_0 j_2  - { 2 \over 5}  \tilde{j}_0  \pa j_2 ] (x) j_2 (x_1) j_2 (x_2) j_2 (x_3) \ra.
}

Now let's take the large $N$ limit in this equation. In the left hand side we can substitute the action of
 the charges on each of the operators. This gives
 \eqn\lehs{
 \la [Q_4, j_2 j_2 j_2]\ra \sim {1 \over \sqrt{N} } \left(\la \partial ^3 j_2 j_2 j_2\ra + \la \partial j_4 j_2 j_2\ra  + {\rm action~on~the~other~}j_2{\rm 's}  \right) }
  Notice that this  is of order  ${1 \over N}$.

In the right hand side of \WIcons\ the  order ${ 1 \over N}$ terms come from
\eqn\termssc{\eqalign{
\int d^3x  ( \pa \la  j_2(x)  j_{2}(x_1)\ra)  \la \tilde{j}_0 (x)   j_2(x_2)  j_2(x_3) \ra \sim \partial^5_1 \int d^3 x
{ 1 \over |x_1 - x| } \la \tilde{j}_0 (x)   j_2(x_2)  j_2(x_3) \ra
}}
where we have integrated by parts and used that all indices are minus ($\partial_1 = \partial_{x_1^-} $).

Now the final result of the integral in the right hand side of \termssc\  is the same as
 the three point function $\la j_0 j_2 j_2 \ra$. Namely, the three point function involving  a twist {\it one} scalar, as opposed to the twist two scalar $\tilde j_0$ that we started with.
  This can be seen by the fact that the integral in \termssc\ (before taking the  $\partial_1^5$ derivative)
   has the same conformal
 properties as $\la j_0 j_2 j_2 \ra$.

  If the current were exactly conserved, we would set \lehs\  to zero. In that case, the only solution is the one corresponding to the free fermion structure in the three point functions.
The reason is that the free boson, or odd solutions of this \Ward identity require a twist one  operator, $j_0$.
Recall that we had said in section six that these three point function ward identities have three independent
solutions involving only the fermion, or only the boson or only the odd structures.
However, in \WIcons\ something remarkable has happened. The right hand side, which comes from the lack of
 conservation of the current, is mocking up perfectly the contribution we would have had in a theory with a
 twist one scalar. This allows for more nontrivial solutions which can be the superposition of all three
 structures for the three point functions.

Notice that if we consider now the \Ward identity
 expanded in the 't Hooft parameter $\lambda$, then  the $\lambda^{0}$ term will satisfy free fermion
result. At the  $\lambda^{1}$ the integral in \termssc\ generates a result equal to the {\it odd} structure
for $\la j_0 j_2 j_2 \ra$. Thus, at order $\lambda^1$,  the odd structure for $\la j_2 j_2 j_2\ra$  is generated.

We expect that further analysis along the lines explained above should fix all the leading $N$ three point functions, exactly
in $\lambda$.

We expect to be able to  apply a similar  ideas to the interacting fixed point of $O(N)$ model.
In this case the spectrum
of leading  $N$ theory is the same, except that the twist two operator is parity even.

Finally, let us comment on the case of weakly coupled gauge theories with matter in the adjoint. In this
case we will typically have single trace operators that can, and will, appear in the right hand side of
the divergence of the currents, once we turn on the coupling. In this case, we also expect to be able to
use the broken \Ward identities to analyze the theory, though it is not clear whether this will be any simpler
than applying usual perturbation theory. One feature is that we will be dealing purely in terms of
observables, without ever discussing things like ``gauge fixing", etc.

\newsec{Case of two conserved spin two currents}

In this section we relax the assumption of a single stress tensor and we generalize the discussion to the
case of two stress tensors. Presumably something similar will hold for more than two, but we have not studied
it in detail.

We now
consider a CFT that has exactly two symmetric traceless conserved spin two currents.
One of them generates conformal transformations and we denote it by $j_2$.
The usual minus generator built out of it is translation along the minus direction $Q_2 = P_{-}$.
Another current we denote as $\hat{j}_{2}$ and the correspondent charge would be $\hat Q_2 = \hat{P}_{-}$.
We assume that these two currents are orthogonal and we normalize their two point functions as
 \eqn\twoptf{
 \la j_{2-} (x_1) j_{2-}(x_2) \ra  = \la \hat{ j}_{2-} (x_1) \hat{j}_{2-}(x_2) \ra = c {(x^+)^{4} \over (x^2)^{d+2} }~,~~~~~~~~~~~  \la  { j}_{2-} (x_1) \hat{j}_{2-}(x_2) \ra =0
 }
 here we used the freedom to rescale $\hat j_2$.

The most general form of the transformation consistent with the conformal properties of generators, two point function \Ward identities and non-zero three point functions is
\eqn\generalB{\eqalign{
[P_{-}, j_2] &=\pa_{-} j_2, ~~~~ [P_{-}, \hat{j}_2] =\pa_{-} \hat{j}_2  \cr
[\hat{P}_{-}, j_2] &= a \,  \pa^2 j_1 + \pa \hat{j}_{2} \cr
[\hat{P}_{-}, \hat{j}_2] &=   \pa j_2 + b \,  \pa \hat{j}_{2}
\cr
}}
We then consider the $\la [\hat{P}_{-}, j_2 j_2 \hat{j}_2] \ra = 0$ \Ward identity, which    sets $a  = 0$.

Now,
for any $b$ we can introduce a new basis
\eqn\newbasis{\eqalign{
J_2 &= {1 \over 2} \left(1 + {b  \over \sqrt{b^2  + 4   }} \right) j_2 - { 1 \over \sqrt{b^2   + 4  }} \hat{j}_2 \cr
\hat{J}_2 &= {1 \over 2} \left(1 - {b  \over \sqrt{b^2   + 4   }} \right) j_2 + { 1 \over \sqrt{b^2  + 4   }} \hat{j}_2
}}
such that the commutation relations take the form
\eqn\generalBnew{\eqalign{
[Q_{J_2}, J_2] &= \pa J_2,~~~~[Q_{\hat{J}_2}, \hat{J}_2] = \pa  \hat{J}_2,  \cr
[Q_{\hat{J}_2}, J_2] &=0,~~~~[Q_{J}, \hat{J}_2] =0.
}}
In this new basis there are two orthogonal $\la J_2 \hat{J}_2 \ra = 0$ conserved spin two currents
such that each of them generates translation for itself and leaves intact the other one.

We can now consider the correlation function with $J_2$'s, $\hat{J}_2$'s and an
insertion of $e^{i Q_{J_2} a}$
\eqn\insertxxx{\eqalign{
\la    \hat{J}_2(y_1) ... \hat{J}_2(y_n)  J_2(x_1) ...  J_2(x_n)   \ra = &\la e^{i Q_{J_2} a}  \hat{J}_2(y_1) ... \hat{J}_2(y_n)  J_2(x_1) ...  J_2(x_n) e^{- i Q_{J_2} a} \ra \cr
  = &
  \la  \hat{J}_2(y_1) ... \hat{J}_2(y_n)   e^{i Q_{J_2} a} J_2(x_1) ...  J_2(x_n) e^{- i Q_{J_2} a} \ra
}}
where we used the fact that $[Q_{\hat{J}_2}, J_2] =0$. Then using the fact that $[Q_{J_2}, P_{-}] = 0$ and also the fact that
$[Q_{J_2}, J_2] = [P_{-}, J_2] $ we can rewrite  \insertxxx\  as
\eqn\decouplingeq{\eqalign{
\la    \hat{J}_2(y_1) ... \hat{J}_2(y_n)  J_2(x_1) ...  J_2(x_n)   \ra = & \la  \hat{J}_2(y_1) ... \hat{J}_2(y_n)   e^{i P_{-} a} J_2(x_1) ...  J_2(x_n) e^{- i P_{-} a} \ra
\cr
= & \la  \hat{J}_2(y_1) ... \hat{J}_2(y_n)    J_2(x_1 + a^{-}) ...  J_2(x_n + a^{-})  \ra.
}}
where all $J_2$'s are translated along the minus direction. By the cluster
decomposition assumption  we get\foot{Let us make this more clear. We first choose the $x_i$ and $y_j$ to be spacelike separated and such that as   $a\to + \infty$  the distances between any   $x_i$ and any  $y_j$ grow.
One can check that such a choice of points is possible, and it still allows us to move all the points in a small
neighborhood without violating this property. Once we establish the
  decoupling for such points, we can analytically  continue the result for all points.
   }
\eqn\decouplingeq{
\la  \hat{J}_2(y_1) ... \hat{J}_2(y_n)  J_2(x_1) ...  J_2(x_n) \ra = \la  \hat{J}_2(y_1) ... \hat{J}_2(y_n) \ra \la    J_2(x_1) ...  J_2(x_n)  \ra.
}
Thus, these correlators are completely decoupled. Notice that, in particular, if we consider
collider physics observables like energy correlation functions for the ``energies'' defined from $J_2$ or $\hat J_2$, we would obtain completely decoupled answers. However, we cannot conclude at this point that we have
two decoupled theories, since we could still consider two theories that share a global symmetry and impose a singlet
constraint on the operators. However, it is clear
that in some sense, the two theories are dynamically decoupled.
All that we have said so far is valid for any theory with two spin two conserved currents, independently of
whether we have any higher spin generator.

\subsec{ Adding higher spin conserved currents}

Now we consider a theory with a conserved higher spin current, $j_s$,
 and exactly two stress tensors.
We know that $[Q_s, j_2 ] \sim j_s$, which implies that $[Q_s , j_s ] = j_2 + \cdots$.
Thus, the right hand side of $[Q_s , j_s]$ has either $J_2$ or $\hat J_2$, or both.
Let us first assume that it has $J_2$. Then the $Q_s$ ward identity on $\la J_2 J_2 j_s \ra$ implies that
there is an infinite number of currents in $ \underline {J_2 J_2 }_b   $ or in $\underline {J_2 J_2 }_f $.
 Now, nothing that appears in the right hand side of these light cone limits can have any non-zero correlator
 with $\hat J_2$, due to \decouplingeq . Thus, from this point onwards, the analysis is effectively the same
 as if we had only one stress tensor.
 Of course, the same holds if $\hat J_2$ appears in the right hand side of $[Q_s , j_s]$.

 One simple example with two conserved spin two currents
  is a free ${\cal N}=1$ supersymmetric theory with $N$
 bosons and $N$ fermions, with an $O(N)$ singlet constraint.
 The two conserved spin two currents are the stress tensor of the boson and the one of the fermions.

The situation is more subtle when we have more than two spin two conserved currents.
 For example, in a theory with two free
bosons, $\phi_1$, $\phi_2$, we have three conserved spin two conserved currents, schematically
  $\pa \phi_1 \pa \phi_1$,
$\pa \phi_1 \pa \phi_2 $ and $\pa \phi_2 \pa \phi_2$. Clearly in this case the theory decouples into two theories,
and not into three!

\newsec{Conclusions and discussion}

 In this article we have studied theories with exactly conserved higher spin currents, with spin $s>2$. We have shown that all correlators of the currents and the stress tensor  are those of a free theory. More precisely, we have made the technical assumption of a single spin two current. Under this assumption, the only two cases are those of a free scalar or a free fermion. More precisely, we proved that the correlators of the currents are the same as in the theory of free bosons or free fermions, but we did
 not demonstrate the existence of a free scalar or a free fermion operator.

This is in the same spirit as the Coleman-Mandula theorem, \refs{\ColemanAD,\HaagQH}, extended here to theories without an
S-matrix.

It can also be viewed as a simple exercise in the bootstrap approach to field theories. One simply starts with the currents, constructs the symmetries, and one ends up fixing all the correlation functions. One never needs to
say what the Lagrangian is. Of course, the answer is very simple, because we end up getting free theories!

In this paper we considered correlators of stress tensors and currents but not of other operators that
the theory can have. In two dimensions, one can find explicitly the correlators of stress tensors, but
that does not mean that the theory is free. On the other hand, in three dimensions we expect that
the simple form of correlators that we obtained will imply that the theory is indeed free. In fact,
one can consider the conformal collider physics observables that come from looking at energies
collected by ``detectors'' at infinity for a state created by the stress tensor \refs{\BashamIQ,\HofmanAR}. The $n$ point energy correlator for a state created by the
stress tensor is schematically $\la 0| T^\dagger(q) \epsilon(\theta_1) \cdots \epsilon(\theta_n) T(q) |0\ra $, where $T(q)$ is the stress tensor insertion with four momentum (roughly) $q$ at
the origin and $\epsilon(\theta)$ are the energies per unit angle
 collected at    ideal calorimeters sitting
 at infinity situated at the angle $\theta$.
These can be
computed by considering $n+2$ correlation functions of stress tensors. Since these agree with
the ones in the free theory, we expect that the theory is free. Notice also that these infrared safe observables are conceptually
rather close to the S-matrix, since they are measured at infinity in Minkowski space.
Nevertheless, one would like to be able to understand directly the constrains of the higher spin
symmetry on other operators and their correlators.  For example, it is natural to conjecture that  their
dimensions are integer or half integer (if the stress tensor is unique). As an example of a theory
that contains an extra operator, we can consider $N$ fermions with an $SO(N)$ singlet constraint
 and the operator
$\epsilon^{i_1, \cdots , i_N} \psi_-^{i_1} \cdots \psi_-^{i_N}$. This can certainly not be produced
from the currents when $N$ is odd. For $N$ odd this has a half integer dimension.

It should be simple to extend these results to four dimensional field theories. In fact, our approach that
uses the light-like limits in section five is expected to work with few modifications. We expect that we will need light-like limits that pick out the free boson, free fermion and free Maxwell fields. In higher dimensions we can have other free fields, such as the self dual tensor in six dimensions.

Recently  there have been many studies of a four dimensional higher spin gravity theory proposed by
Vasiliev (\VasilievBA\ and references therein) . These theories have higher spin {\it gauge} symmetry in the bulk. They also have an
$AdS_4$ vacuum solution. If one chooses AdS boundary conditions that preserve the higher spin symmetry, then our methods show that the theory is essentially the same as  a free theory on the boundary.
Thus, we have proven the conjecture in \KlebanovJA\ for the free case.   More
concretely it is the theory of $N$ free bosons or free fermions with an $O(N)$ singlet constraint. Given that the free case works, then \GiombiYA\ showed that the conjecture
 followed for the interacting case. Other arguments were presented in \refs{\KochCY,\DouglasRC}.
      If the
boundary conditions break the higher symmetry in a slight way, then one can also use the symmetries
to constrain the results. Notice also that our results for the quantization of $\tilde N$
  show that the coupling constant
in  a unitary  Vasiliev theory with higher spin symmetric boundary condition is quantized.

It is also interesting that theories with almost conserved higher spin currents can be analyzed in this spirit. We discussed some simple computations in the case of large $N$ vector models (recently studied in \GiombiKC,\AharonyJZ ).
Here one can use the slightly broken symmetries to get interesting results about the correlators. Of course, our previous analysis of the exact case is the backbone of the analysis for the slightly broken one. We suspect that correlators in these theories can be completely fixed by these considerations alone, and it would be nice to
carry this out explicitly.

Notice that for any weakly coupled theory we have a slightly broken higher spin symmetry.
In general, the breaking of the higher spin symmetry can occur at the single trace level. This occurs, for
 example, in gauge theories with adjoint fields. One might be able to perform perturbative
 computations    using the higher spin currents and
the pattern of symmetry breaking. Of course, one would recover  the  results of
standard perturbation theory. However, the fact that we work exclusively with gauge invariant physical
observables might lead to important simplifications, particularly in the case of
 gauge theories. Notice, by the way,   that  the
bilocal operators that we have introduced in \bilocsp\ \conffu\ are simply
 the (Fourier transform) of the   operators
whose matrix elements are the parton distribution functions.
In an interacting theory, these operators also have
a Wilson line connecting the two ``partons''.

More ambitiously, one would like to understand large $N$ limits of theories with adjoints, such as
${\cal N}=4$ super Yang Mills, in the 't Hooft limit.  In general, the single trace
higher spin currents can  have
large anomalous dimensions compared to one.
 However, the anomalous dimensions are still small compared to $N$.
It is natural to  wonder whether they continue to impose interesting constraints.
This is closely related to the possible emergence of a useful higher spin symmetry in
string theory at high energies (see e.g. \SagnottiAT\ and references therein).

Interestingly, one expects
that the {\it absence} of   single trace higher spin states in the low energy spectrum,
namely the fact that all single trace higher  spin operators have large anomalous dimensions,
 should impose an intriguing constraint: The theory has an AdS dual well approximated by Einstein { gravity}.
  For more discussion of this point see  \refs{\HeemskerkPN,\FitzpatrickZM,\papadodimas}.

\vskip .3in \noindent

{\bf Acknowledgments}

  We would like to thank N. Arkani-Hamed, T. Dumitrescu, A. Dymarsky,  T. Hartman, A. Jevicki, G. Pimentel, S. Rychkov,
 N. Seiberg, X. Yin,  E. Witten for discussions.

This work   was  supported in part by U.S.~Department of Energy
grant \#DE-FG02-90ER40542. The work of AZ was supported in part by the US National Science Foundation under Grant No.~PHY-0756966 and by the Department of Energy under Grant No.\#DE-FG02-91ER40671.

\appendix{A}{ Argument for $[Q_s , j_2] = \partial j_s + \cdots$  }

We now argue that the action of $Q_s$ on the stress tensor produces $j_s$ in the right hand
side, where $j_s$ is the current we used to construct $Q_s$,
\eqn\stress{
[Q_{s}, j_2 ] = \pa j_{s} + ... ~,~~~~~~~~~~~~~~~~s>1
}
This follows from the fact that we know how the charge $Q_{s}$ transforms under the conformal group.

 The general form of the commutator is
\eqn\fixed{
[Q_{s}, T_{\mu \nu}] = c   \partial_- J_{ --\cdots -  \mu \nu} + ....
}
Here we eliminated a possible term like $ \partial_\mu J_{\nu -\cdots - }$ by
applying $\partial_\mu$ to both sides of \fixed\ and noticing that the left hand side, as well
as the right hand side of \fixed\ are zero, but the extra possible term would not be zero.
Now, let us
prove that $c$  is not zero.
Imagine it is zero,  then contract \fixed\
 with dilaton Killing vector and integrate over $d^{d-1} x$ to get
\eqn\fixedB{
[Q_{\zeta}, \int d^{d-1} x T_{0 \nu} x^{\nu}] = [Q_{\zeta}, D] = 0
}
which is inconsistent with the fact that $[Q_{\zeta}, D] = (s-1) Q_{\zeta}$.
Note that for $s=1$, the current does not appear in the right hand side. As usual the
stress tensor should be invariant under global symmetries.


\appendix{B}{An integral expression for the odd three point functions }

\ifig\OddIntegral{ Constructing the three dimensional correlator as arising from a
four dimensional integral in a CFT. The three dimensional space lies at $t=0$. } {\epsfxsize2.2in\epsfbox{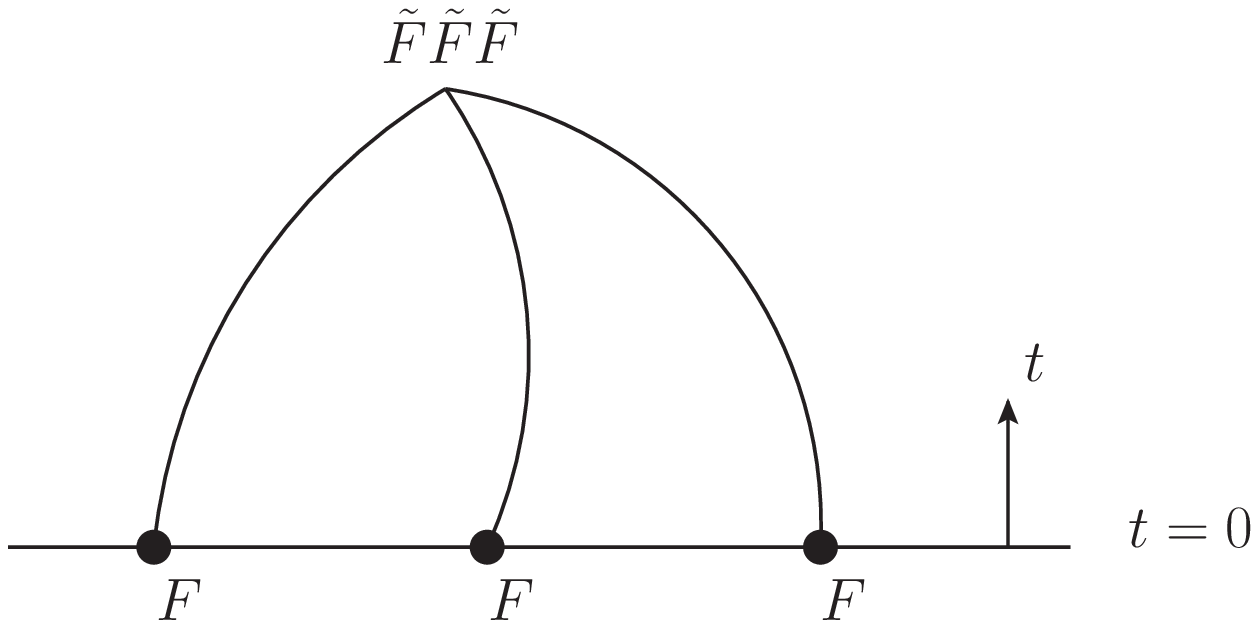}}

In this appendix we provide a
``constructive'' way to write it and a nice rationale for its existence.
The idea is inspired by the $AdS$ expressions for the odd pieces for spin one and spin two
currents discussed in \MaldacenaNZ .  However, we will use something even simpler. We   view the three
point function as a correlator on a domain wall defect inside a (non-unitary)
four dimensional conformal field theory.
We construct it as follows. We start with four dimensional operators with  spins $(j_l,j_r)= ( s,0), ~ (0,s)$ at the four dimensional unitarity bound.
We then introduce a quartic bulk interaction that preserves conformal symmetry. We then consider the correlator
of three of these fields on a three dimensional subspace $t=0$, to leading order in the interaction, see \OddIntegral .

We denote the  four dimensional operators with spins
$(j_l,j_r)= ( s,0), ~ (0,s)$ by  $ F_{\alpha_1 ... \alpha_{2 s} }  , ~  \tilde{F}_{\dot{\alpha}_1 ... \dot{\alpha}_{2 s}}$. Of course,
these operators are at the four dimensional unitarity bound and given in terms of  free fields.
The conformal dimension is   $[F] = s+1$ which is the same as dimension of conserved currents in $d=3$ CFT.
Their correlators are
\eqn\propself{
\la F_{\alpha_1 ... \alpha_{2 s}} (x_1)  \tilde{F}_{\dot{\alpha}_1 ... \dot{\alpha}_{2 s}} (x_2) \ra = { \prod_{i=1}^{2 s} (x_{12})_{\alpha_{i} \dot{\alpha}_{i}} \over (x_{12}^2)^{2 s+1}}
}
we can contract indices with polarization spinors $\lambda^{\alpha}$ and $\bar \mu^{\dot{\alpha}}$ so that we get
\eqn\strspinB{
\la F( x_1, \lambda )  \tilde{F} ( x_2, \bar \mu)  \ra = { (\lambda  {\slash\!\!\! x_{12} } \bar \mu)^{2 s} \over (x_{12}^2)^{2 s+1}} = (\lambda  {\slash\!\!\! \partial }_{x_{12}} \bar \mu)^{2 s} { 1 \over x^2_{12} } .
}

Now let's consider the self dual interaction of the form
\eqn\inter{
 g \int  d t d^3 \vec{z} \chi  \tilde{F}_{s_1} \tilde{F}_{s_2} \tilde{F}_{s_3}
}
where all indices are contracted using $\epsilon^{\dot{\alpha} \dot{\beta}}$.
Here $\chi$ is a scalar  operator of dimension $[\chi] = 1 - (s_1 + s_2 + s_3) $.
The contraction of indices is possible  when the triangle inequality is satisfied
\eqn\trianglein{
s_{i} \leq s_{i+1} + s_{i+2}, \quad  i=1,2,3 \ \ {\rm mod} \ 3
}
 This interaction
breaks parity in the bulk so that we can expect that it will generate parity breaking structure for the correlators at $t=0$, where $t$ is the fourth coordinate.

Now imagine we are computing   correlation functions of three $F_{\alpha_1 \cdots \alpha_{2s}}$ inserted at
the boundary $t=0$. There is a subset of four dimensional conformal transformations which
map this boundary to itself. They act on this boundary as the three dimensional conformal
transformations. From that point of view the $F_{\alpha_1 \cdots \alpha_{2 s} }$ operators transform as
three dimensional twist one operators of spin $s$. The contraction with the spinor $\lambda_{\alpha}$ gives us
the same as contracting three dimensional currents with the three dimensional spinors, as in
\GiombiRZ .
In addition, one can see that they are conserved currents from the
three dimensional point of view\foot{This needs to be checked separately because the interaction \inter\
is not obviously unitary and, thus, we cannot use the unitarity bound argument.}. For this purpose it is enough to show that
 the propagator is annihilated by  three dimensional
  divergence operator $\pa_{\lambda} {\slash\!\!\! \partial_{x}} \pa_{\lambda}$.
The structure of \strspinB\ shows that it is enough to check it for $s=1$. For $s=1$ the
current is just $j_i =
  F_{ti} +   \epsilon_{i j k }F_{jk}$. Then the divergence is zero due to
  the Maxwell equation and the Bianchi identity.

We now postulate that $\chi$ in \inter\ has expectation value $\langle \chi \rangle = t^{-\Delta_{\chi}} = t^{ s_1 + s_2 + s_3 -1}$. This expectation value is consistent
with the three dimensional conformal invariance, and can be viewed as arising from a
``domain wall defect'' at $t=0$.

We can now consider the expression that results from considering the three point correlator at first order
in the interaction \inter
\eqn\diagrambulk{\eqalign{
&\la F_{s_1}(\vec{x}_1, \lambda_1) F_{s_2}(\vec{x}_2, \lambda_2) F_{s_3}(\vec{x}_3, \lambda_3) \ra \sim \int_{0}^{\infty} d t d^3 \vec{x}_0 t^{s_1 + s_2 + s_3 - 1} \cr
&{(\lambda_1 {\slash\!\!\! x}_{10} {\slash\!\!\! x}_{02} \lambda_2)^{ (s_1 + s_2 - s_3)} (\lambda_1 {\slash\!\!\! x}_{10} {\slash\!\!\! x}_{03} \lambda_3)^{ (s_1 + s_3 - s_2)} (\lambda_2 {\slash\!\!\! x}_{20} {\slash\!\!\! x}_{03} \lambda_3)^{(s_2 + s_3 - s1)} \over (t^2 + (\vec{x}_1 - \vec{x}_0)^2)^{2 s_1 + 1} (t^2 + (\vec{x}_2 - \vec{x}_0)^2)^{2 s_2 + 1} (t^2 + (\vec{x}_3 - \vec{x}_0)^2)^{2 s_3 + 1}}
}}
The operators in the left hand side are inserted at $t=0$. They can be viewed as three dimensional conserved currents.

This is a conformal invariant expression for three dimensional correlators of conserved currents.
It  contains both even and odd contributions.

Notice that factors in the  numerator of \diagrambulk\ have the schematic form
\eqn\denfactor{
(\lambda_1 {\slash\!\!\! x}_{i 0} {\slash\!\!\! x}_{0j} \lambda_2) \sim  t (\vec{n} \vec{x}_{12}) + \eps_{i j k} n^{i} x_{01}^{j} x_{02}^{k}
}
where $\vec{n} = \lambda_1^{\alpha} \vec{\sigma}_{\alpha \beta} \lambda_2^{\beta} =\lambda_2^{\alpha} \vec{\sigma}_{\alpha \beta} \lambda_1^{\beta} $.
We see that these
two pieces behave differently under three dimensional parity.
Thus, to extract the odd piece we can subtract parity
conjugated integral,  which can also be extracted by changing $t\to -t$.
 In the end one can show that the parity odd piece is given by extending the range
 of integration
\eqn\oddgen{\eqalign{
&\la j_{s_1}(\vec{x}_1, \lambda_1) j_{s_2}(\vec{x}_2, \lambda_2) j_{s_3}(\vec{x}_3, \lambda_3) \ra_{odd} \sim \int_{-\infty}^{\infty}  d t d^3 \vec{x}_0 t^{s_1+s_2+s_3-1} \cr
&{(\lambda_1 {\slash\!\!\! x}_{10} {\slash\!\!\! x}_{02} \lambda_2)^{ (s_1 + s_2 - s_3)} (\lambda_1 {\slash\!\!\! x}_{10} {\slash\!\!\! x}_{03} \lambda_3)^{ (s_1 + s_3 - s_2)} (\lambda_2 {\slash\!\!\! x}_{20} {\slash\!\!\! x}_{03} \lambda_3)^{(s_2 + s_3 - s_1)} \over (x_{10}^2)^{2 s_1 + 1} (x_{20}^2)^{2 s_2 + 1} (x_{30}^2)^{2 s_3 + 1}}.
}}
This formula is valid both for integer and half integer spins.
If $s_1 = s_2 = s$, then one can check that under the exchange $(\lambda_1 ,x_1) \leftrightarrow (\lambda_2 ,x_2) $
we get a factor of $(-1)^{s_3 + 2 s}$. This implies that the correlator for two identical currents
$\la j_s j_s j_{s_3} \ra $ is zero if $s_3$ is odd. Note that this is also zero when $s$ is half integer, since in this case we expect an anticommuting result when we exchange the first
two currents.

\subsec{Light cone limit for the odd three point function}

For the odd piece  the light cone limit give us
\eqn\lightodd{
 \langle \underline {j_{s_1} j_{s_2} }_o j_{s_3} \rangle  = \partial^{s_1}_1  \partial^{s_2}_2 \Upsilon(s_3), ~~~~s_3 > 0
}
where
\eqn\basic{\eqalign{
\Upsilon(s) &= {1 \over \hat{x}_{12} }  \left[{\hat{x}_{12} \over \hat{x}_{13} \hat{x}_{23} } \right]^{s} \sum_{k=0}^{[{s - 1 \over 2}]} \pmatrix{s-k-1 \cr k}{1 \over 2 k + 1}  \left[{\hat{x}_{13} \hat{x}_{23} \over \hat{x}_{12}^2} \right]^{k}
}}
here $[\alpha]$ means integer part of $\alpha$ and $\hat{x}$ were defined in \hatx . And $\pmatrix{n \cr m}$ is the usual binomial coefficient. Interestingly, in this limit the answer can be written explicitly for any $s$.
Notice that we do not have any square roots of the $x_{ij}$'s. Essentially for the same
reason, we will not have a $|y_{12}|$ as we the limit $y_{12} \to 0$, we only get $1/y_{12}$.
In general, it can be checked from the expressions in \GiombiRZ\  that all square roots
disappear from the odd correlator, even away from the light cone limit.

\appendix{C}{No odd piece from the one point energy correlator}

Let's consider the case when $\la j_2 j_2 j_2\ra$ is the boson piece plus the
odd  piece. In other words, $\la j_2 j_2 j_2 \ra_f =0$.

Consider the one point energy correlator for the state created by a stress tensor at
the origin   with  energy $q^0$ and zero spatial momentum. It is characterized by a
two dimensional polarization tensor $ \eps^{i j} T_{i j}$. See \HofmanAR\ for more details.
The most general $O(2)$ symmetric structure in this case is given by the following expression
\eqn\fourone{\eqalign{
\la {\cal E}(\vec{n}) \ra &=  {q^0 \over 2 \pi} \left[1 +  t_4({ |\eps_{i j} n^{j} n^{j}|^2 \over  \eps_{i j} \eps^{*}_{i j} } - {1 \over 4} ) \right] \cr
&-{q^0 \over 2 \pi}  d_4 \left[ {\varepsilon^{i j} ( n_{i}  n^{m} \eps_{j m} \eps^{*}_{k p}  n^{k}  n^{p} +   n_{i}  n^{m} \eps^{*}_{j m} \eps_{k p}  n^{k}  n^{p}) \over  2  \eps_{i j} \eps^{*}_{i j}}\right],
}}
it's even part was studied in \BuchelSK , however, the odd piece was missed in the literature before as far as we know.
The positive energy condition $\la {\cal E}(\vec{n})  \ra \geq 0$ becomes
\eqn\condfour{\eqalign{
|d_{4}| \leq \sqrt{16 - t_4^2}.
}}

For the case of the boson it is easy to check that $t_4 = 4$.
 Thus, we are forced to set $d_4 = 0$.
We got that in the theory of Majorana fermion $t_4 = -4$.

More intuitively consider the state created by
$T_{11} - T_{22}$. Then the one point energy correlator is given by
\eqn\onepconc{
\la {\cal E}(\vec{n}) \ra =  {q^0 \over 2 \pi} \left[1 +  {t_4 \over 4} \cos 4 \theta  + {d_4 \over 4} \sin 4 \theta \right]
}
where $\theta$ is the angle between $x^1$ and the detector.

For a free boson, we have that the energy correlator vanishes at $\theta = \pi/4$ which forces
$t_4 = 4~$  \foot{The operator is odd under the exchange of the 1 and 2 axis, but the state of two
massless scalars going back to back along this axis    would be even. }.
 Then near $\theta = \pi/4$ we would have regions with negative energy correlator
if $d_4$ was nonzero.

Thus, we are forced to  conclude that the odd piece is not allowed by the assumption of the positivity of the energy flow.
 In the main text we have also excluded the odd piece in other ways.

Notice also that the energy correlator that we obtain, after setting the fermion and odd parts to zero, is such
that it vanishes for one particular angle.
 This is already a suggestion that the theory is probably free, since
in an interacting theory we expect showering so the energy
distribution will never be exactly zero. In other words,
showering from a neighboring angle would make it non-zero.
 In fact, it is natural to conjecture that if  $\la j_2 j_2 j_2 \ra_f =0$ (or $\la j_2 j_2 j_2 \ra_b =0$), then the
 theory is free, without assuming that a higher spin current exist.

\appendix{D}{ Half-integer higher spin currents implies even spin higher spin currents }

Here we assume that we start from a higher, {\it half}-integer spin current $j_s$, $s\geq 5/2$.
We then argue that we also  have  even spin currents. Then we can go back to the   case treated in the main text.

The analysis is completely parallel to what was done in section five for integer spin currents.
 We simply need a
couple of other formulas.
For half integer $s$ and $s'$  we need that
\eqn\eqnsli{ \eqalign{
& \langle \underline { j_{s'} j_2 }_b j_s \ra \propto  \partial_1^{s'-1/2} \partial_2^2 \la
\underline{ \psi \phi }  j_s \ra_{free} ~,~
\cr
& \langle \underline { j_{s'} j_2 }_f j_s \ra \propto   \partial_1^{s'-1/2} \partial_2  \la
\underline{ \phi \psi }  j_s \ra_{free} ~,~
\cr
& \la  \underline{ \psi \phi }  j_s \ra_{free}  \propto
 { 1 \over \hat{x}_{13}^{3/2} \hat{x}_{23}^{1/2} } \left( { 1 \over \hat{x}_{13} } -
{ 1 \over \hat{x}_{23} } \right)^{s-1/2}
\cr
& \la  \underline{ \phi \psi }  j_s \ra_{free}  \propto
 { 1 \over \hat{x}_{13}^{1/2} \hat{x}_{23}^{3/2} } \left( { 1 \over \hat{x}_{13} } -
{ 1 \over \hat{x}_{23} } \right)^{s-1/2}}}
It is clear that in a free  theory of bosons and fermions
 these formulas are true.  In fact, these formulas follow from conformal invariance and current conservation
 of the third current. This can   be shown directly using the methods of appendix I.

We can now use these expressions to consider the $Q_s$ \Ward identity on
$\la \underline {j_2 j_2}_b j_s \ra$. $[Q_s , j_s]$ gives a sum of integer spin currents so
that we get a contribution to the \Ward identity similar to the second term in \wards .
The action of $[Q_s , \underline {j_2 j_2}_b ]$ requires a bit more words. First $[Q_s , j_2]$
gives some other half integer spin currents. Then we can do the light-like OPE using \eqnsli\
for each term and summing all the terms. After the dust settles, such terms give us
\eqn\actiqs{ \eqalign{
 \langle [Q_s , \underline {  j_2 j_2}_b ] s \rangle = &  \partial^2_1 \partial_2^2 \left( \gamma \partial_1^{s-{ 3\over 2}}   { 1 \over \hat{x}_{13} } + \delta   \partial_2^{ s -{ 3 \over 2 } } { 1 \over { \hat{x}_{23} }} \right)  { 1 \over \sqrt{ \hat{x}_{12} \hat{x}_{13} }} \left( { 1 \over \hat{x}_{13} } - { 1 \over \hat{x}_{23} } \right)^{s- { 1 \over 2}}
 \cr
 \langle [Q_s , \underline {  j_2 j_2}_f ] s \rangle = &  \partial_1 \partial_2
  \left( \gamma \partial_1^{s- {1\over 2} }  { 1 \over \hat{x}_{23}  } + \delta   \partial_2^{ s - {1 \over 2}  } { 1 \over \hat{x}_{13} }  \right)
  { 1 \over \sqrt{ \hat{x}_{12} \hat{x}_{13} }  }
 \left( { 1 \over \hat{x}_{13} } - { 1 \over \hat{x}_{23} } \right)^{s-{1\over 2} }
 }}
here permutation symmetry fixes $\delta = (-1)^{s - {1 \over 2}} \gamma $.

 The analysis of these \Ward identities is very similar to the one we did before.
 Again we can count the number of independent terms in the right hand side and
 conclude that we expect a unique solution.
 This unique solution is the one we get in a
 free supersymmetric theory. We conclude that currents $j_k$
 appear in $\underline { j_2 j_2}$ with even spins $k = 2, 4, \cdots , 2 s -1$.
  If $s\geq 5/2$, then  $ 2 s -1\geq 4$ and we have a higher spin current with integer
  spin and we can go back to the analysis done in  section five.

 \appendix{E}{Functions that are annihilated by all charges }

Suppose that we have a  function $f(x_i)$  of $n$ variables that has a Taylor
 series expansion and is such that $Q_s = \sum_i ( \partial_{x_i} )^s$
  annihilates it for all odd $s$.
 We want to prove that    the function can
 be written as a sum of functions (not all equal)
 \eqn\conclu{
 f(x_i ) = \sum_{\sigma} g_\sigma (x_{\sigma(1)} -x_{\sigma(2)}, x_{\sigma(3)} - x_{\sigma(4)} ,..., x_{\sigma(n-1) } - x_{\sigma(n ) })
 } where the sum is over all permutations. Notice that the each function depends on
  $[{n \over 2}]$ variables,  compared to the original $n$ $x_{i}$.
 If $n$ is odd,   then we drop the last variable  (the function is
 independent of the last variable). Here we will not use conformal symmetry. Also we assume
 that the variables $x_i$ are ordinary numbers and not vectors, etc.

 \subsec{Proof}

 First if $f$ has a Taylor series expansion, we can organize it in terms of the overall
 degree of each term. Say we have a polynomial in $x_i$ and we can separate the terms
 according to the overall degree of each term. Then $Q_s$ should annihilate terms of
 different degrees. In other words, there is no mixing between terms of different degrees.
 Thus, we have to prove the statement for polynomials of degree $k$.
 The statement is obvious for degree $k=0$. Now assume it is valid up to degree $k$ and
 we want to prove it for degree $k+1$.

 If we have a polynomial $P_{k+1}$ that is annihilated by the charges, then we can take
 $\partial_{x_1} P_{k}$ which is also annihilated and is a polynomial of degree $k$.
 Thus, by assumption $\partial_{x_1} P_k $ can be written as in \conclu .
 We can now integrate each term in such a way that we preserve its form. For example
 if we have a function of the form $g(x_1 - x_2 , \cdots)$, then we integrate it with
 respect to $x_{12}$. After we are done, we would have written
 $P_{k+1} = \sum_\sigma g_\sigma + h(x_2, \cdots x_n)$. In other words, we are left
 with an integration ``constant'' which is a function of $n-1$ variables. Clearly the first
 term, the one involving the functions $g$, is annihilated by the $Q_s$, so $h$ is also
 annihilated. Now we can repeat the argument for $h$, viewing it as a function of $n-1$ variables. In this way we eliminate all variables.

 Thus, we have proven what we wanted to prove.

 \subsec{Functions with Fourier Transform}

 Of course, if the function $f$ has a Fourier transform, then it is even easier to prove the
 statement. In that case we have $Q_s =\sum_i k_i^s$. Now we take $s\to \infty$, then the
 largest or lowest value of $k$ dominates (depending on whether there is a $|k|$ bigger than
 one
 or not). There must be an even number of largest $|k_i|$ otherwise for large $s$ we would
 violate the equality. Say there are two, say $k_1 = -k_2$. Thus, the Fourier transform must
 have a $\delta(k_1 + k_2)$. We can now repeat the same with the second largest, etc, to argue
 that $f = \sum_\sigma   g( k_{\sigma(2)}, k_{\sigma(4)}, \cdots , k_{\sigma(n) } )
 \delta( k_{\sigma(1) } + k_{\sigma(2) } ) \cdots \delta( k_{\sigma( n-1)} - k_{\sigma(n) } ) $. which is the same as saying \conclu .

 The reason we proved it for functions that have a Taylor series expansion, rather than a Fourier expansion,
 is that our definition of the quasi-bilocal operators,  \correc\ and  \correcf , involve the operator
 product expansion. So these are given in terms of power series expansions. In principle, we do not know how
 to continue them beyond their convergence radius. For this reason we do not know, in principle, whether a
 Fourier transform exists, or whether the action of the higher spin charges \propeqn\ continues to be valid
 beyond that region. Of course, the result proved in this appendix, together with conformal symmetry and analyticity, allows us to extend the bilocals everywhere.

 \appendix{F}{Current conservation equation in terms of cross ratios}

We consider parity even correlation functions of a conserved current with
 two spinning operators of general twist
 \eqn\correlf{
  \langle  O_{\tau_1 , s_1} (x_1)   O_{\tau_2, s_2}(x_2) j_{s_3}(x_3) \rangle= { 1 \over |x_{12}|^{\tau_1 + \tau_2 - 1} |x_{13}|^{\tau_1 + 1 - \tau_2} | x_{23}|^{\tau_2 + 1 - \tau_1 } }F( Q_i , P_i )
  }
  where $Q_i$ and $P_i$ are the conformal cross ratios introduced in \GiombiRZ, see also
  \CostaMG .
  Here $\tau_{2,3}$
  are the twists of the second and third operators.
  The function $F$ is constrained by current conservation.
  Here we give an expression for the condition for current conservation as a
  function of the cross ratios. The idea is to write $\pa_{\lambda_3} {\slash\!\!\! \partial_{x_3}} \pa_{\lambda_3}$, act on \correlf ,  and then rewrite the answer in terms of
  cross ratios. We end up with an equation for $F$ which can be expressed purely in terms
  of a differential operator acting on $F$, ${\cal D}_3 F =0$, with

  \eqn\expro{ \eqalign{
  {\cal D}_3 =
     &    -
  8   ( Q_2 Q_3 - P_1^2 ) \partial_{Q_2} \partial^2_{Q_3}    +
  8   ( Q_1 Q_3 - P_2^2 ) \partial_{Q_1} \partial^2_{Q_3}
  \cr
  &
  - 4 Q_2 \partial_{Q_2} \partial_{Q_3} + 4 Q_1 \partial_{Q_1} \partial_{Q_3}
  - Q_1 \partial_{P_2}^2   +Q_2 \partial_{P_1}^2 +
  \cr
  &+ 2( - Q_1 P_1 + P_2 P_3) \partial_{P_1} \partial^2_{P_2} -
  2( - Q_2 P_2 + P_1 P_3) \partial_{P_2} \partial^2_{P_1} + \cr
  &
  + 2 ( - Q_1 Q_2 + P_3^2 )(  \partial_{P_2}^2 \partial_{Q_2} -\partial_{P_1}^2 \partial_{Q_1} )
  +
  \cr
  &+ 2 ( P_2^2 - Q_1 Q_3 ) \partial_{Q_3} \partial_{P_2}^2
  -  2 ( P_1^2 - Q_2 Q_3 ) \partial_{Q_3} \partial_{P_1}^2 +
  \cr
 &+  8 ( Q_1 P_1 - P_2 P_3) \partial_{Q_1} \partial_{Q_3} \partial_{P_1}
 - 8 ( Q_2 P_2 - P_1 P_3) \partial_{Q_2} \partial_{Q_3} \partial_{P_2} +
 \cr &+   (\tau_2 -\tau_1) \left[ - 4 \partial_{Q_3}  Q_3 \partial_{Q_3}  -
     4 \partial_{Q_3} ( P_1 \partial_{ P_1}  + P_2 \partial_{P_2} )
     - Q_2 \partial^2_{P_1} - Q_1 \partial^2_{P_2}  - 2 P_3 \partial_{P_1} \partial_{P_2} \right] \cr}}

In principle one could also write a similar expression for odd structures. One can write the
$S_i$ in \GiombiRZ , by taking square roots in (2.18) of \GiombiRZ ,
and then being careful about the
sign of the square root. Doing this, we can still use \expro , for odd structures.

\appendix{G}{Four point function of scalar operators in the fermion-like theory}

Using the transformation law for the twist two scalar which we derived in section 6.6
\eqn\findaB{
[ Q_4 , \tilde j_0 ] = \partial^3 \tilde j_0  + \gamma (\partial_-^2  j_{2 - \perp} - \partial_-
\partial_{\perp} j_{2--} )
}
we can solve the \Ward identity $\la [Q_4,\tilde{j}_0 \tilde{j}_0 \tilde{j}_0 \tilde{j}_0] \ra = 0$.
The result is
\eqn\fourscalarFer{\eqalign{
\la  \tilde j_0  \tilde j_0  \tilde j_0  \tilde j_0  \ra &=  {f (u,v) \over x_{12}^4 x_{34}^4} \cr
f  (u,v) &= f  (v,u) = {1 \over v} f ({u \over v}, {1 \over v}) \cr
f  (u,v) &= \alpha (1 + {1 \over u^2} + {1 \over v^2}) + \beta  {1 +u^{5/2}+v^{5/2} - u^{3/2}(1+v) - v^{3/2}(1+u) - u -v  \over u^{3/2} v^{3/2}}
}}
and for  $\la J_2 \tilde{j}_0 \tilde{j}_0 \tilde{j}_0 \ra$ we get,
using the notation of \CostaMG ,
\eqn\oddfermionstr{\eqalign{
\la J_2 (x_1)   j_0 (x_2)  j_0 (x_3)  j_0 (x_4) \ra &= {\eps (Z_1, P_1, P_2, P_3, P_4) \over (P_1 P_2)^3 (P_1 P_3) (P_1 P_4) (P_3 P_4)^{2}} \cr &\left[Q_{123} f_1 (u,v) + Q_{134} f_2 (u,v) + Q_{142} f_3 (u,v)  \right], \cr
Q_{1 i j } &= {(Z_1 P_i) (P_j P_1) - (Z_1 P_j) (P_i P_1) \over (P_{i} P_{j})}, \cr
\gamma f_1 (u,v) &= {9 \over 5} \beta {v^{3/2} \over u^{3/2}}, ~\gamma f_2 (u,v) = {9 \over 5} \beta {u \over v^{3/2}},~\gamma f_3 (u,v) = {9 \over 5} \beta {1 \over u^{3/2} v^{3/2}}.
}}
As in the case of bosons $\beta$ and $\gamma$ are fixed as soon as we choose the normalization for the two-point function of stress tensors.

\appendix{H}{Conformal blocks for free scalar}

In the main text we introduced the quasi-bilocal operators by taking a particular light-like limit of stress tensors \correc . These are the sum of contributions from several spins.
They can be defined in any theory, even in theories that do not have the higher spin
symmetry. In a generic interacting theory we only get the contribution of the stress tensor.
Here we would like
to present some explicit formulas showing that the projection on to a single spin contribution
does not define a   genuine bilocal operator.
%

 Let us focus on the contribution with a particular spin $s$. This is again a
  quasi-bilocal operator  which   can be defined as follows
\eqn\defbiloc{\eqalign{
b_{s} (   x_1,x_2   ) &= \sum c_{i,n} (x_1-x_2)^i \pa^n j_{s} ({x_1+x_2 \over 2}), \cr
\la b_{s} (   x_1,x_2   )  j_{s}(x_3) \ra &\sim \la   \phi^{*}(x_1) \phi(x_2)  j_{s} (x_3)\ra_{free}.
}}
The second line fixes the constants in the first line. We can define it even away from the light cone limit by
this formula.
Defined in this way,  $ b_{s} (  x_1,x_2   )$ transforms as a  bi-primary operator with conformal weights $(\Delta_x = {1 \over 2}, \Delta_y = {1 \over 2})$
and obeys the Laplace equation for both points.
 However, as we emphasized,  this is not the product of two free fields!.
  In particular, notice that $b_2 (x_1,x_2)$ exists in any CFT.

To see the quasi-bilocal nature of $b_s (x_1,x_2)$ let us  consider the two point function of bilocals $\la b_s (x_1, x_2) b_s (x_3, x_4) \ra$. In $d=3$ it is given by the contribution of conserved current of spin $s$, $J_s$, into the four point function of free fields. This problem was addressed recently in
\DolanDV . The answer is given by the formula (6.20) in \DolanDV\ with $a=b=0$, $\lambda_1 = s + {1 \over 2}$, $\lambda_2 = {1 \over 2}$, $l=s$

\eqn\confblock{\eqalign{
\la b_s (x_1, x_2) b_s (x_3, x_4) \ra &= {1 \over |x_{12}|  |x_{34}|} {\cal F}_{s} (u, v) \cr
 {\cal F}_{s} (u, v) &= { \sqrt{u} \over \pi} \int_{0}^{\pi} d \theta {4^{s} X^{s} (1 + \sqrt{1- X})^{-2 s} \over  \sqrt{1- X}} \cr
X &= z \cos^2 \theta + \bar{z} \sin^2 \theta \cr
u &= z \bar{z}, ~~~~ v = (1 - z) (1 - \bar{z})
}}
This integral is not known for general $s$. Fortunately,  for $s=0$ it is known and we get
\eqn\confblockZero{\eqalign{
\la b_0 (x_1, x_2) b_0 (x_3, x_4) \ra &= {1 \over |x_{13}|  |x_{24}|} {2 \over \pi}  {K (- {z - \bar{z} \over 1 - z}) \over \sqrt{1 - z}}
}}
where $K(y)$ is the complete elliptic integral of the first kind.

One can check that this solution satisfies the Laplace equation. Also one can check that it has a singularity at the expected locations.   However, the behavior near
the singularity is not the one that is expected for the correlator of local operators. More precisely, in the limit 
$z \to 1$, with $\bar z$ fixed but close to one, we get a term $\log(1-z)$.
For   local operators the singularities are power-like.

\appendix{I}{Fixing correlators in the light-like limit}

Here we would like to show how the three point
correlators are fixed in the light-like limit $ \la \underline{j_{s_1} j_{s_2}} j_{s_3} \ra$.
We use the conventions and notation of Giombi, Prakash  and Yin \GiombiRZ .
In the light like limit we have $P_{3} =0$, the rest of the conformal invariants are non-zero.

The general conformal invariant, without any factor of $P_{3}$ has the form
\eqn\confi{
\la j_{s_1} j_{s_2} j_{s_3} \ra = { 1 \over |x_{12}| |x_{13}| |x_{23}| }
\sum_{a, b} c_{ab} Q_1^a Q_2^b P_{2}^{ 2 s_1 - 2 a } P_{1}^{ 2 s_2 - 2 b }
Q_3^{ s_3 - s_1 - s_2 + a + b }.
}
We can now use the relation  the $P$'s and $Q$'s given in equation (2.14) of \GiombiRZ .
Setting $P_{3}=0$ we get
\eqn\getrel{
P^2_{1} Q_1 + P_{2}^2 Q_2 = Q_1 Q_2 Q_3 ~,~~~~~ \Rightarrow  Q_3 = { P_{1}^2 \over Q_2 } + { P_{2}^2 \over Q_1 }
}
We can use this relation to solve for $Q_3$ and eliminate it from \confi . Then we find a simpler formula
\eqn\solcan{\eqalign{
\la j_{s_1} j_{s_2} j_{s_3} \ra = & { 1 \over |x_{12}| |x_{13}| |x_{23}| } f(s_1,s_2,s_3)~,~~~~
\cr
f(s_1,s_2,s_3) = & \sum_{a=0}^{s_3} c(a) Q_1^{ s_1 -a} P_{2}^{ 2 a } P_{1}^{  2 (s_3 -a ) } Q_2^{ s_2 - s_3 + a }
}}

We would like now to impose current conservation on the third current. The correspondent operator of the divergence can be obtained
from \expro\ by setting $P_{3}$ to zero. The resulting operator is

\eqn\operac{
{\cal D }_3 = - ( 1 + 2 P_1 \partial_{P_1} + 2 Q_2 \partial_{Q_2} ) Q_1 \partial_{P_2}^2 + (1 + 2 P_2  \partial_{P_2} + 2 Q_1 \partial_{Q_1} ) Q_2 \partial^2_{P_1}.
}

By imposing ${\cal D }_3  f(s_1,s_2,s_3) = 0$ we  find a recursive relation
\eqn\recrel{
{ c(a+1) \over c(a) } = {  ( s_1 + a + {1 \over 2}  ) \over  ( a + {1 \over 2} )} { ( s_3 - a  - {1 \over 2} )    \over   ( s_3 +  s_2 -  a - {1 \over 2}) } {( s_3 -a ) \over ( a+1 ) }
}

This gives a unique solution, which is the one  we considered in the main text \limitbcorr .
Once we know the solution is unique, we can simply use the free boson answer to get \limitbcorr .
But one can check explicitly that \recrel\ gives \limitbcorr .

\subsec{Fermions}

For fermions we expect a factor of $P_{3}$. In whatever multiplies this factor, we can set $P_{3}=0$ and
use \getrel . This means that the conservation condition will act in
the same way on what multiplies $P_{3}$ as it acted on the boson case.
Thus, we get the following structure
\eqn\somthli{
  f^f(s_1,s_2,s_3) = P_{3} P_{1} P_{2} \sum_{b=0}^{\tilde s_3} \tilde c(b)   Q_1^{ \tilde s_1 -b} P_{2}^{ 2 b } P_{1}^{  2 (\tilde s_3 -b ) } Q_2^{ \tilde s_2 - \tilde s_3 + b }
}
 where $\tilde s_i = s_i -1$.This looks very similar to \solcan . In fact,
 we see that
\eqn\ferco{
f^f(s_1,s_2 ,s_3) = P_{3} f_{h}( s_1 -{ 1 \over 2 },s_2 -{ 1 \over 2 }, s_3 )
}
where $f_{h}$ is the same sum as in \solcan\ but with $a$ running over half integer values
$a ={ 1 \over 2}, {3 \over 2} , \cdots , s_3- { 1 \over 2 } $.
Thus, the action of ${\cal D}_3$, \operac ,  leads to the same recursion relation \recrel , but now
running over half integer values of $a$.

This again proves that there is a unique structure, which is \limitfcorr .

In the same way one can obtain \eqnsli .

\subsec{General operators}

We now explicitly solve the problem of finding the different parity even structures we can have for
the three point function of two operators of the same twist and one conserved current. The problem is very
similar to what we solved above. The main observation is that $P_{3}$ does not depend on $\lambda_3$ or $x_3$.
Then if we have a solution to the current conservation condition, we can generate a new solution by
multiplying by $P_{3}$.
Then the general even structures have the form
\eqn\genstr{
\langle O_{s_1} O_{s_2} j_{s_3} \rangle \sim  { 1 \over |x_{12} |^{ 2 \tau_0 -1}  |x_{23}| |x_{13}|}   \sum_l   P_{3}^{2 l }  \left[
\langle j_{s_1-l} j_{s_2 -l } j_{s_3} \rangle_b + \langle j_{s_1-l} j_{s_2 -l} j_{s_3} \rangle_f  \right]
}
The sum runs over the bosonic and fermionic three point functions of three conserved currents. We can easily
check that these are solutions. The fact that these are all the solutions is obtained by taking light cone limits.
To order $P_{3}^0$ the solution is constrained as in \recrel , which is the same as what we get from the $l=0$
in \genstr . If we now subtract the $l=0$ term, which is a solution, we can take the light cone limit and focus
on the $P_{3}^1 $ term, etc.

In \genstr ,
 the sum over the bosonic structures runs over $ 1 + {\rm min} [s_1,s_2]$ values. The sum over
fermion structures runs over $ {\rm min} [s_1,s_2]$ values.
The total number of structures that we have is
\eqn\numbestr{
 1 + 2 \,  {\rm min} [ s_1 , s_2 ]
}
These formulas are valid for integer spins.
Probably, there is a similar story with half integer spins.

It would be interesting to fix also the structures that appear when the twists of the two operators are not
the same. For that we would have to include the last term in \expro\ when we analyze the constraints.

\appendix{J}{ More explicit solutions to the \Ward identities of section five  }

Here we discuss further the \Ward identities of section five and we prove that they
have the properties stated there.
To find the explicit solution of \wards\ it is convenient to restate the problem in a slightly different language.
First remember that it is easy to prove that there is a unique solution of \wards\ which coincides with the free field one.
However, we need to check  that for  all even $k$  $\tilde{\alpha}_k$ is
 non-zero. So we focus on the free field theory.

\subsec{ Expression for the currents in the free theory }

The expressions for the conserved currents are bilinear in the fields and contain a
combination of derivatives that makes the field a primary conformal field. These
are computed in multiple places in the literature, (see e.g. \MikhailovBP\ \KonsteinBI\ and
references therein).
We would like to start by providing a derivation for the expression for these currents
that is particularly simple in three dimensions.

Consider the Fourier space expression for a
current contracted with a three dimensional spinor $\lambda$,
$ \lambda^{2s}. J(q)$, where $q^\mu$ is the three-momentum.
We can  consider the matrix element
\eqn\matrel{
 F = \la p_1, p_2 | (\lambda^{2s }. J(q)) |0\ra
 }
 here $p_1$ and $p_2$ are two massless three dimensional momenta of the two bosons or two
 fermions that make up the current. A massless three dimensional momentum can be written in
 terms of a three dimensional spinor, $\pi_\alpha$. In other words, we write
  $p_{\alpha \beta} = p^\mu
 \sigma^\mu_{\alpha \beta} = \pi_{\alpha } \pi_{\beta}$. Since we have two massless momenta
 we have two such spinors $\pi^1$ and $\pi^2$. Of course, $q = p_1 + p_2$ and it is not an independent variable.  We can also view \matrel\ as the form factor
 for the current.
  It is telling us how the current is made out of the fundamental fields.
 Clearly,  Lorentz invariance implies that
 \matrel\ is a function $F( \lambda . \pi^1 , \lambda . \pi^2 )$. However,
 since the current is also conserved, it should obey the equation
 \eqn\equcu{
 0=   q^{\alpha \beta} { \partial^2 \over \partial \lambda^\alpha \lambda^\beta } F =
 (\pi_1 . \pi_2 )^2 \left[{ \partial^2 \over \partial  ( \pi^1.\lambda)^2} +
{ \partial^2 \over \partial  ( \pi^2.\lambda)^2} \right] F
 }
 where we expressed $q_{\alpha \beta}  = \pi_\alpha^1\pi_\beta^1 + \pi_\alpha ^2 \pi_\beta^2 $
 and we acted on the function $F( \lambda . \pi^1 , \lambda . \pi^2 )$.
We can now define the variables
\eqn\varia{
z  = \pi_1 . \lambda  - i \pi_2 . \lambda ~,~~~~~~~~~~ \bar z = \pi_2 . \lambda + i \pi_2 . \lambda
}
Then the equation \equcu\ becomes
\eqn\newh{
\partial_z \partial_{\bar z} F(z,\bar z) = 0
}
The solutions are purely holomorphic or purely anti-holomorphic functions. Since
the homogeneity degree in $\lambda$ is fixed to be $2 s$, we get the two solutions
\eqn\finso{
 F_b = z^{2s} + { \bar z }^{ 2 s } ~,~~~~~~~~~~F_{f} =  z^{2s} - { \bar z }^{ 2 s }
}
The first corresponds to the expression for the current in the free boson theory and the
second to the expression in the free fermion theory, as we explain below. These expressions
contain the same information as the functions $\la \phi \phi^* j_s \ra_{free}$ and $\la \psi \psi^* j_s \ra_{free} $ that we discussed in section five.

In order to translate to the usual expressions in terms of derivatives acting on fields,
without loss of generality,
we can   choose $\lambda$ so that it is purely along the direction $\lambda^-$.
In the boson expression, each $\pi^i.\lambda$ appears
 to   even powers, say $(\pi^1.\lambda)^{2 k} (\pi^2 . \lambda)^{2 s - 2 k} $ such  a term can be replaced by
$\partial_-^k \phi^* \partial_-^{2 (s-k)} \phi $. The $F_b$ gives
 the expression
 \eqn\conscurf{
j^{bos}_{s} \sim \sum_{k=0}^{s} (-1)^{k}  {1 \over  (2 k)! ( 2 s - 2 k)!}  \pa^k \phi^{*} \pa^{s-k} \phi.
}
 Similarly, for the fermion case we have odd powers and we replace
  $(\pi^1.\lambda)^{2 k+1} (\pi^2 . \lambda)^{2 s - 2 k -1} $ by
  $\partial_-^{ k } \psi^*_- \partial_-^{ s - 1 - k} \psi_-$ and $F_f $ in \finso\ gives
%
  \eqn\conscur{
j^{fer}_{s} \sim \sum_{k=0}^{s} (-1)^{k}  {1 \over (2k+1) ! (2 s - 2 k -1)! }
  \pa^k \psi^{*} \pa^{s-k} \psi.
}
where all are minus component.

Notice that we do not even need to go to Fourier space to think about the $z$ and $\bar z$
variables. We can just think of them as book-keeping devices to summarize the expansions
\conscur\ , \conscurf\ via the simple expressions \finso .

\subsec{Analyzing the \Ward identity}

Symmetries act on free field as
\eqn\symfree{
[Q_{s}, \phi] = \pa^{s-1} \phi, \quad [Q_{s}, \phi^{*}] =(-1)^{s} \pa^{s-1} \phi^{*}
}
and a similar expression for the fermions. Expressing this in terms of the $z$ and $\bar z$
variables we rewrite \symfree\ as
\eqn\actgenpol{
[Q_s, j_{s'}] (z, \bar{z}) =  \left[ (z+\bar{z})^{2 s-2} -  (z-\bar{z})^{2 s-2} \right](z^{2 s'} \pm \bar{z}^{2 s'}) .
}
where the $\pm$ indicate the results in a boson of fermion theory respectively.
Then, using the notation  from \wards,  we have
\eqn\tocheck{
[Q_s, j_s] (z, \bar{z}) = \sum_{k = 1}^{s-1} \tilde{\alpha}_{2 k} z^{2s -2k -1} \bar{z}^{2 s - 2 k -1}  (z^{4 k} \pm \bar{z}^{4 k})
}
where we used the map $\pa_3 \to z \bar z$. As in the previous subsection this
 can be seen by $\partial_3 = \partial_{x_3^-} \to \lambda^2 q$ and $\lambda^2 .q = ( \lambda . \pi^1)^2 + ( \lambda . \pi^2)^2 = z \bar z$.  Expanding the left hand side of \actgenpol\ for
 $s'=s$ we get
\eqn\solalpha{
 \tilde{\alpha}_{2 k} = { 2 \ \Gamma (2 s -1) \over \Gamma (2 k) \Gamma (2 s - 2 k)}
}
which are all non-zero. Thus, this solves the problem of showing that $\tilde \alpha_k$ are
non-zero for the free theories.
This works equally simply for the boson-like and the fermion-like
light like limits.

We can also  do a similar expansion for the action of $[Q_s , j_{s'}]$ and
find which terms appear in the right hand side
 \eqn\tocheck{\eqalign{
[Q_s, j_{s'}]  = &  ( 2 s-2 )! \sum_{r = -s'-s+2}^{s+s'-2} \tilde \alpha_r \partial ^{ s+s'-r -1}j_{r}
\cr
 \tilde \alpha_r = &  {[1 + (-1)^{s+s'+r}] \over 2 } \left( { 1 \over \Gamma( r+ s - s') \Gamma(s+s'  -r) } \pm { 1 \over \Gamma(r  + s + s') \Gamma(s-s' -r) }\right).
 }}
The $\pm$ indicates the boson/fermion case.

This is how all the other \Ward identities
we discussed in section five can be analyzed. Namely, in section five we argued that
all solutions are unique. And by computing for free theories as in this appendix
we check which coefficients
are non-zero. This works in the same way for the boson-like and fermion-like light like limits.

We can go over the \Ward identities and the  properties we used.

\wardint :
We used that if $s$ is odd, $s>1$, and $s'=2$, then  both  $j_1$ and $j_{s}$ appear
 in the right hand side.
This simply amounts to checking that for $s'=2$, $r=1, s$, the coefficient in \tocheck\ is
non-zero for any odd $s$. We can also check that we get a $j_s$ in the right hand side.
This is the case both for the boson and fermion case.

\wardfi :
We used that if $s$ is even and $s'=2$, then we get  both a  $j_2$ and a $j_s$ in the
right hand side of \tocheck .

\wardi :
 Here we used a slightly different property. The reason is that
\wardi\ does not correspond to the transformation of any free field, since an $s$ even
charge in free theories acts with $+$ signs in front of the $\partial^{s-1}$.
In order to analyze \wardi , it is convenient to view the functions in \wardi\ in
Fourier space, as we have done here. Then we can rewrite \wardi\ as
\eqn\wardinew{
 \left[ (z+\bar{z})^{2 s-2} +  (z-\bar{z})^{2 s-2} \right](z^{2 } \pm \bar{z}^{2 }) = (2 s - 2)!
 \sum_{r = -s}^{s} \tilde \beta_r  (z \bar z)^{ s -r} ( z^{2 r} \pm {\bar z}^{ 2 r } )
}
Then one can find that
\eqn\find{
\tilde\beta_r =  {[1 + (-1)^{s + r}] \over 2} \left( { 1 \over \Gamma( r+ s - 1) \Gamma(s  -r + 1) } \pm { 1 \over \Gamma(r + s + 1) \Gamma(s -r-1) }\right)
}
which is the same expression as before, in \tocheck , for $s'=1$, except for
the $(-1)$ dependent prefactor which now implies that
  now the sum over $r$
runs from $r = -s, -s + 2 , \cdots , s$.
Now the important point is that $\tilde \beta_s$ is non-zero. We used this in \wardi .

In the discussion of elimination of $j'_0$, for the boson-like case, we used that
$[Q_s , j_{s'}]$ produces both $j_0$ and $j_2$ in the boson theory and for $s$, $s'$ even, with
$ s>s'$. This can also be checked from \tocheck , by setting $r=0,2$. Note that in \tocheck\
$\tilde \alpha_0 =0$ in   fermion theory (the minus sign in \tocheck), as expected.

\listrefs
\bye